\documentclass{aa}    

\usepackage{graphicx}
\usepackage{txfonts}
\usepackage[colorlinks=true, citecolor=blue, linkcolor=black, urlcolor=black]{hyperref}

\usepackage{natbib}
\usepackage{amssymb}
\usepackage{color}

\newcommand{\dd}{\mathrm{d}}
\newcommand{\thetabar}{\bar\theta}

\begin{document} 

  \title{The spin state of asteroid Apophis and a prediction of its change during the 2029 close encounter with Earth}
  
  \titlerunning{The spin state of asteroid Apophis and the prediction of its change}

  \author{J.~\v{D}urech		  \inst{1}    \and
          D.~Vokrouhlick\'y     \inst{1}    \and  
          P.~Pravec             \inst{2}    \and
          K.~Hornoch            \inst{2}    \and
          P.~Ku\v{s}nir\'ak     \inst{2}    \and
          P.~Fatka              \inst{2}    \and
          H.~Ku\v{c}\'akov\'a   \inst{1,2}  \and
          J.~Hanu\v{s}		  \inst{1}    \and
          M.~Ferrais            \inst{3}     \and 
          E.~Jehin              \inst{4}     \and 
          Z.~Benkhaldoun        \inst{5,16}     \and 
          O.~Humes              \inst{6}     \and 
          D.~Polishook          \inst{7}     \and
          M.~Marsset            \inst{8}     \and           
          G.~McMillan           \inst{9}     \and           
	       E.~Podlewska-Gaca     \inst{10}     \and
    	     M.~Colazo             \inst{10}     \and
    	     A.~Marciniak          \inst{10}     \and
    	     K.~Kami\'nski         \inst{10}     \and
    	     M.~K.~Kami\'nska       \inst{10}     \and
    	     S.~Zo\l{}a            \inst{11}     \and
    	     M.~Dr\'o\.zd\.z       \inst{12}     \and
    	     W.~Og\l{}oza          \inst{12}     \and
    	     M.~\.{Z}ejmo          \inst{13}     \and
    	     B.~Carry              \inst{14}    \and
          D.~E.~Reichart         \inst{15}}

  \institute{Charles University, Faculty of Mathematics and Physics, Institute of Astronomy, V Hole\v{s}ovi\v{c}k\'ach 2, 180\,00 Prague, Czech Republic,
             \email{durech@sirrah.troja.mff.cuni.cz} \and
            Astronomical Institute of the Czech Academy of Sciences, Fričova 298, 251 65 Ond\v{r}ejov, Czech Republic \and
            Florida Space Institute, University of Central Florida, 12354 Research Parkway, Partnership 1 building, Orlando, FL, 32828, USA \and 
            Space sciences, Technologies and Astrophysics Research Institute, Universit\'e de Li\`ege, All\'ee du 6 Ao\'ut 17, 4000 Li\`ege, Belgium \and 
            Department of Applied Physics and Astronomy, and Sharjah Space and Astronomy Hub, University of Sharjah, United Arab Emirates \and 
            Technische Universitaet Braunschweig, Institute of Geophysics and Extraterrestrial Physics, Braunschweig, Germany\and 
            Faculty of Physics, Weizmann Institute of Science, Rehovot 0076100, Israel \and 
            European Southern Observatory, Alonso de Córdova 3107, Vitacura, Casilla 19001, Santiago, Chile. \and 
            Department of Earth, Atmospheric, and Planetary Sciences, Massachusetts Institute of Technology, 77 Massachusetts Avenue, Cambridge, MA 02139, USA \and 
            Astronomical Observatory Institute, Faculty of Physics and Astronomy, Adam Mickiewicz University, S{\l}oneczna 36, 60-286 Pozna\'n, Poland \and 
            Astronomical Observatory, Jagiellonian University, ul. Orla 171, 30-244 Krakow, Poland \and 
            Mt. Suhora Astronomical Observatory, University of the National Education Commission, ul. Podchorazych 2, 30-084 Krakow, Poland \and 
            Janusz Gil Institute of Astronomy, University of Zielona Góra, Lubuska 2, 65-265 Zielona Góra, Poland \and 
            Université Côte d’Azur, CNRS-Lagrange, Observatoire de la Côte d’Azur, CS 34229, 06304 Nice Cedex 4, France \and
            Department of Physics and Astronomy, University of North Carolina at Chapel Hill, Chapel Hill, NC 27599, USA \and 
            Oukaimeden Observatory, High Energy Physics, Astrophysics and Geoscience Laboratory, Faculté des Sciences Semlalia (FSSM), Cadi Ayyad University, Marrakesh, Morocco
         }

  \date{Received ?; accepted ?}

  \abstract
  {On April 13, 2029, the asteroid Apophis will pass near Earth at a geocentric distance of about $38,000$~km. This will provide a unique opportunity to study the effects of Earth's gravitational torque on the asteroid's spin state and figure. Numerical models have suggested that the post-encounter spin state will critically depend on the orientation of Apophis during the flyby.}
  {We aim to determine the spin state of Apophis from its photometric observations collected during two apparitions in 2012--2013 and 2020--2021. This will enable us to accurately predict the pre-encounter rotation state and, by accounting for Earth's gravitational torque, predict a range of possible post-encounter states.}
  {We used the light curve inversion method for tumbling asteroids to reconstruct the spin state of Apophis and its convex shape model. The result is adopted as the initial condition of a numerical model describing Apophis's future rotation state.}
  {The data from the two apparitions are insufficient to determine Apophis's rotation and precession periods uniquely. The formally best-fit solution is $P_\phi = 27.374 \pm 0.001$\,h for the precession period and $P_\psi = 262.2 \pm 0.1$\,h for the rotation period, but at least two other combinations of the periods provide a similarly good fit to the available data. All the currently acceptable models result in approximately the same pre-encounter orientation of Apophis in early 2029 (within $20^\circ$ in terms of Euler angles). This is because the accurate photometric data were collected during two apparitions separated by 8 years, which is the same interval as from 2021 to 2029. Although the close encounter with Earth in April 2029 hugely increases the post-encounter uncertainty of Apophis's spin state, the short-axis spin mode will be preserved with a high likelihood.}
  {Additional observations taken in 2027 and 2028 will break the ambiguity in Apophis's pre-encounter spin solution and allow us to get a more accurate post-encounter spin state prediction.}


  \maketitle

  \nolinenumbers
  
  \section{Introduction}

    The asteroid (99942)~Apophis is a near-Earth object classified as a potentially hazardous asteroid. It received significant attention after reaching the Torino scale of 4 for a few days in late 2004. The probability of Apophis impacting Earth in April 2029 quickly dropped with further observations, but its very close approach to Earth, which is predicted to occur on April 13, 2029, is still a rare event for a body of this size \citep{Bro.ea:26}. Physically, Apophis is characterized as an Sq-class asteroid \citep{Bin.ea:09} with a mean diameter of about 340--420\,m \citep{2014A&A...566A..22M,Lic.ea:16,Bro.ea:18,Bro.ea:26}, slowly rotating in an excited rotation state \citep{Pra.ea:14, Lee.ea:22}. A significant amount of effort has been devoted to precisely determining its orbit and future impact probabilities \citep{Far.ea:13b, Vok.apophis:15}. 

    Its upcoming encounter with Earth on April 13, 2029, presents a great scientific opportunity to study this object and promote the importance of planetary defense \citep{2022PSJ.....3..123R}. One key scientific aspect of the flyby is that Earth's gravitational torques will alter Apophis's rotational state \citep{2023Icar..39015324B}. The exact outcome depends on Apophis's attitude during the flyby and its tensor of inertia. Possible tidally driven surface movement depends on the interior and regolith structure \citep{Yu.ea:14}. In general, this unique natural experiment can be used to extract information about Apophis's interior that would otherwise be inaccessible. There are at least three space missions planned to explore Apophis: OSIRIS-APEX, the NASA mission that will start proximity operation after the encounter \citep{2023PSJ.....4..198D, Nol.ea:25}; the ESA mission RAMSES that will rendezvous with Apophis even before it passes Earth \citep{Laz.ea:25,Mic.ea:25}; and DESTINY+, the JAXA mission that is planned to fly by Apophis in February 2029 \citep{Ara.ea:25}. Although Apophis will be studied in detail before and after it encounters Earth, to maximize the scientific output from the post-encounter observations, Apophis's pre-encounter properties must be known. In particular, its spin state will change during its flyby towards Earth, so the information about the pre-encounter spin state is critical to any interpretation of how the spin would be affected. Moreover, knowing the spin state precisely will enable us to predict the flyby attitude, compute the effects of torques, and indicate the post-encounter spin state, which is essential for proximity operations of spacecraft.

    This paper aims to use photometric data of Apophis (Section~\ref{sec:data}) to reconstruct its shape and spin model (Section~\ref{sec:model}). From this model, we predicted its post-encounter spin state (Section~\ref{sec:encounter}). 

  \section{Photometric data}
  \label{sec:data}

    Apophis was discovered in June 2004. The first photometric data came from January 2005 (V.~Reddy) and are available at the Asteroid Lightcurve Photometry Database.\footnote{\url{https://alcdef.org}} However, the data do not contain much information about Apophis's rotation, as they are only differential magnitudes and cover only three nights.

    Other data from the same apparition are presented on the web page of R.~Behrend\footnote{\url{https://www.astro.unige.ch/~behrend/page_cou.html}}. They are also only relative; see the discussion by \cite{Pra.ea:14}.

    During the next favorable apparition of Apophis in 2012--2013, \cite{Pra.ea:14} took an extensive photometric dataset covering an interval from December 2012 to April 2013 and found that Apophis is an excited (tumbling) spin state. They constructed a convex shape model of Apophis, with the precession period $P_\phi = 27.38 \pm 0.02$\,h and the rotation period $P_\psi = 263 \pm 2$\,h in the \cite{Kaa:01} convention. This model agreed with radar delay-Doppler observations from the same apparition. From the radar observations, a few possible non-convex shape models were constructed \citep{Bro.ea:18}.

        Many observers have used the latest observing window of Apophis in 2020--2021 to obtain more photometric data. We conducted observations with the 1.54m Danish telescope at the La Silla Observatory in Chile from November 16, 2020, to May 6, 2021. The telescope (IAU/MPC code W74) is equipped with a $2048 \times 2048$~px CCD detector with a pixel size of $13.5\,\mu$m.  This configuration provides a plate scale of 0.40~arcsec~pixel$^{-1}$ and a field of view of $13.5 \times 13.5$~arcmin$^{2}$.  We observed Apophis on 67 individual nights, collected 1280 data points calibrated in the Cousins R photometric system, covering a phase angle range of 24--100$^\circ$ (Table~\ref{tab:aspect_P2021}). 
        The telescope was tracked at half of the apparent sky motion rate of the asteroid.  We processed and photometrically reduced the observations using our standard procedure as described in \cite{Pra.ea:14}, \cite{Fat.ea:25}, and the references therein.  The absolute accuracy of the Cousins R calibrations using the \cite{Lan:92} standard stars was 0.01\,mag, which makes it a unique dataset when compared with other photometric light curve datasets that we used as relative data, as described below.

        We initially adopted Transiting Exoplanet Survey Satellite \citep[TESS;][]{Ricker2015} photometric data of Apophis from Sector 35, spanning February 19 to March 7, 2021, as reported by \citet{Lee.ea:22} and processed following the scheme of \citet{Pal2020}. While these data were incorporated into our analysis, our initial shape modeling revealed significant inconsistencies between the best-fitting solution and the TESS light curves, in contrast to the good agreement obtained with other datasets. To investigate this issue, we independently extracted the TESS photometry using the pipeline developed by \citet{Humes2024}. Apophis was observed with two detectors (CCD 1 of Camera 1 and CCD 1 of Camera 2). The Camera 2 dataset contains a $\sim$4-day gap, during which we discarded anomalous points that showed imperfect background subtraction. According to the Sector 35 release notes\footnote{\url{https://archive.stsci.edu/missions/tess/doc/tess_drn/tess_sector_35_drn51_v02.pdf}} \citep{Fausnaugh2021}, TESS lost fine pointing during this interval. Since our reduction does not include an additional alignment step on the background-subtracted frames, the affected measurements are unreliable and were excluded from further analysis. The dataset obtained using the \citet{Humes2024} pipeline performed significantly better in shape modeling (Sec.~\ref{sec:multiple_solutions}), and we therefore adopted it in place of the data from \citet{Lee.ea:22}.
                
        We used time-series photometry of Apophis with the two TRAPPIST 0.6-m robotic telescopes -- TRAPPIST-South at ESO La Silla, Chile (TS) and TRAPPIST-North at the Oukaimeden Observatory, Morocco (TN) -- on multiple nights between January 15 and April 13, 2021 (see Table~\ref{tab:aspect}). The twin Ritchey–Chrétien systems are operated remotely from Liège and designed for high-precision CCD photometry. All observations were conducted with the wide Exo filter, and the instrumental magnitudes were calibrated against field stars to the Cousins R system. For technical descriptions of the facilities, see \citet{Jehin2011}.
        
        We utilized photometric observations of Apophis from \citet{2022PSJ.....3..123R} obtained with the 0.7-m telescope at the Wise Observatory, Israel, over 15 nights between December~24, 2020, and April~7, 2021. All images were acquired using a wide Luminance filter, and the photometry from each night was relatively calibrated against nearby field stars \citep{Polishook2020}.
        
        Furthermore, we obtained lightcurve data of Apophis during five nights at the Observatoire du Mont-Mégantic (OMM) in Québec, Canada. The observatory operates a 1.6-m Ritchey–Chrétien telescope equipped with two CCD cameras. We used the PESTO instrument with the Sloan photometric filters.

        We also utilized photometric data collected by \cite{Lee.ea:22} from various sources, kindly provided by the first author. Their dataset consists of more than 200 light curves observed with 36 telescopes. 

        Other photometric data were obtained by observers participating in the Gaia-GOSA project \citep{Santana-Ros2016} and by individual professional observers. In particular, we used 
        (i)~the 1.54 m reflecting telescope located at the Estación Astrofísica de Bosque Alegre  (EABA, MPC code 821) situated in the Sierras Chicas, Córdoba, Argentina; 
        (ii)~the 0.7m Roman Baranowski Telescope (RBT/PST2) of the Adam Mickiewicz University located at the Winer Observatory in Arizona, USA; 
        (iii)~the Panchromatic Robotic Optical Monitoring and Polarimetry Telescopes (PROMPT) built by the University of North Carolina at Cerro Tololo Inter-American Observatory (CTIO) in Chile \citep{Zola2021};
        (iv)~the 0.6m Cassegrain telescope at the Mount Suhora Observatory in Poland;
        (v)~the 0.6m telescope at the Adiyaman University Application and Research Center in Adiyaman, Turkey; and 
        (vi)~the 0.5m OAUJ-CDK500 telescope \citep{Zola2025} of the Astronomical Observatory of the Jagiellonian University.
        The data are listed in Table~\ref{tab:aspect}.

  \section{Determining the spin state}
  \label{sec:model}

    Because of Apophis's excited rotation, the shape and spin state reconstruction from photometric data is more complicated than in the case of a principal axis rotator. The method of light curve inversion for tumbling asteroids was developed by \cite{Kaa:01}, and we used the same formalism here. The spin state is described by the angular momentum vector $\vec{L}$; initial orientation defined by three Euler angles $(\phi_0, \theta_0, \psi_0)$ of precession, nutation, and rotation, respectively; and normalized principal axes $I_1, I_2$ of the inertia tensor (with $I_3 = 1$). For practical reasons of the inversion method, the four parameters $\vec{L}$ and $\theta_0$ were substituted with an equivalent set of parameters that are the direction of the angular momentum vector in ecliptic coordinates $(\lambda, \beta)$, the rotation period $P_\psi$, and the precession period $P_\phi$. The details and other relevant references can be found in the Appendix of \cite{Kaa:01}. The asteroid's shape was modeled as a convex polyhedron with facet areas approximated by a series of harmonic functions. Coefficients of the series thus represent shape model parameters \citep{Kaa.Tor:01}. The order and degree of the harmonic expansion were eight, so the total number of shape parameters was 80 because the zero-order parameter corresponds to the size, which is not constrained in our model. The algorithm starts from initial parameter values and converges to a local minimum in the standard $\chi^2$ metric.
    
    This method of light curve inversion was already used on Apophis by \cite{Pra.ea:14} and \cite{Lee.ea:22}. The model by \cite{Pra.ea:14} was constructed from only one apparition data, so the uncertainties of the determined spin parameters, namely the rotation and precession periods, were relatively large and do not allow predictions of Apophis's orientation in 2029 to be made. To accurately predict the orientation of Apophis in 2029, we needed to know the periods precisely; therefore, using data from both apparitions, 2012--2013 and 2020--2021, was necessary.  The formal uncertainties of the model constructed by \cite{Lee.ea:22} are small enough for extrapolation to 2029, but, as we demonstrate below, there are also other spin solutions present that provide the same or even better fits than the model made by \cite{Lee.ea:22}. In this context, analyzing the uniqueness of the light curve inversion solution and providing a realistic estimation of the model uncertainties is critical.

    The available photometric dataset is huge and heterogeneous. To speed up the computations and ensure that systematic errors would not affect our solution, we selected only the most accurate data subset: the data taken with the 1.54m Danish telescope (DK154) in 2012--2013 (158 points) and 2020--2021 (1280 points), and the TESS data (1149 points). For the 2020--2021 apparition, the DK154 dataset is the most homogeneous, and it covers the widest interval of geometries (Table~\ref{tab:aspect_P2021}), while the TESS data almost continuously cover the longest time interval of about 15 days. 

    Although the data from different observatories vary in their photometric accuracy, we did not take this into account and used all data points with the same weights. This simplified approach is justified because the introduced error is much smaller than the errors caused by systematic effects and the ambiguity in the spin solution. Different weights would affect the model parameters on the same order as the difference between models A and A* described in the next section.

    \subsection{Multiple solutions}
    \label{sec:multiple_solutions}
\begin{figure*}[t]
        \includegraphics[width=\textwidth]{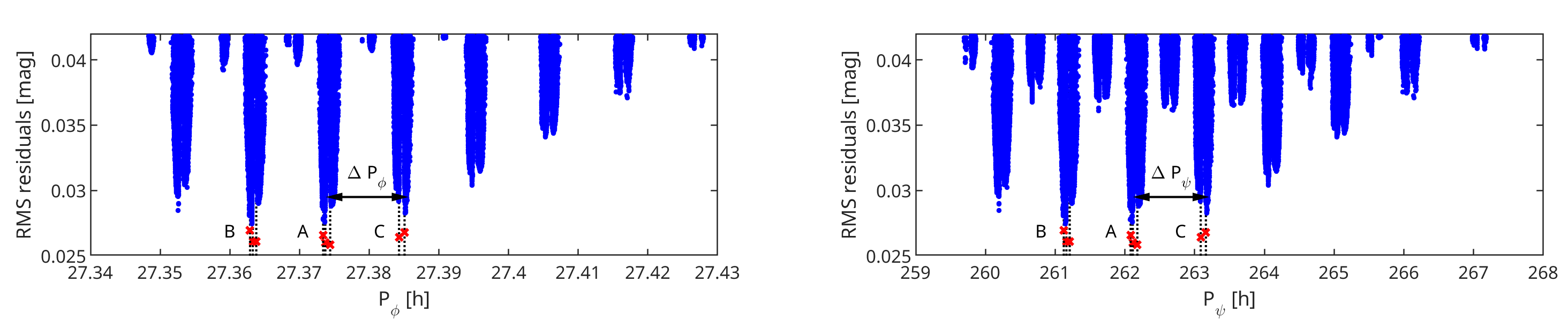}    
        \caption{Period scan. The plots show the residuals (blue points) of the fit for different values of the precession period $P_\phi$ ({\it left\/}) and the rotation period $P_\psi$ ({\it right\/}). The dotted vertical lines indicate the positions of the eight best solutions, and the red crosses represent RMS residuals for the high-resolution models. The distances between local minima $\Delta P_\phi$ and $\Delta P_\psi$ are denoted.} 
        \label{fig:period_scan}
        \end{figure*}

\begin{figure}
        \includegraphics[width=\columnwidth, trim=3.5cm 12cm 19cm 1cm, clip]{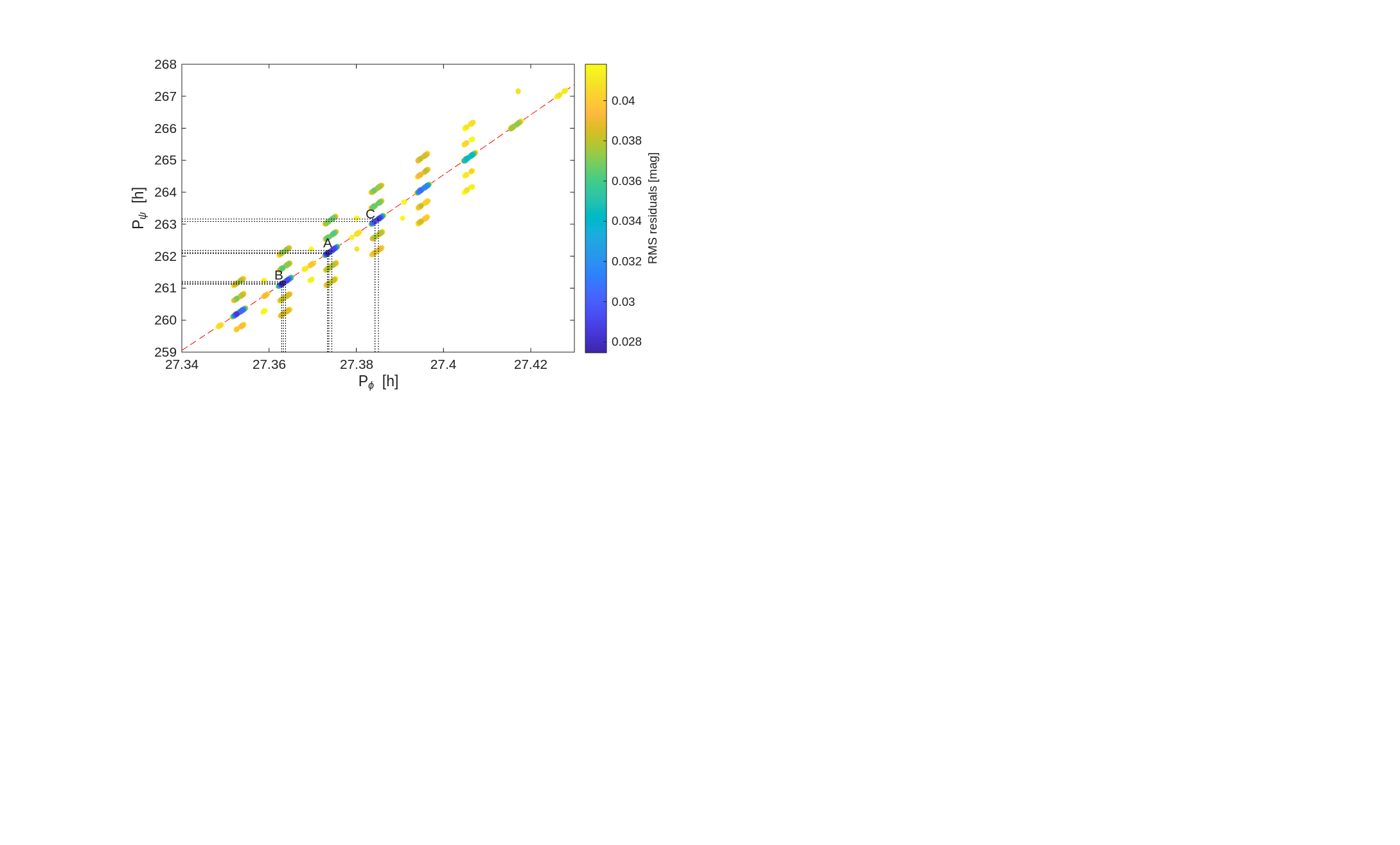}    
        \caption{Period scan. The plot shows the same data as in Fig.~\ref{fig:period_scan}, but here the RMS residuals are color coded, and both the precession period $P_\phi$ and rotation period $P_\psi$ are plotted to visualize the correlation between them. The best solutions with the lowest residuals (blue color) lie on the red dashed line that is in fact part of a hyperbola satisfying the relation $(P_\phi^{-1} - P_\psi^{-1})^{-1} = 30.5658$\,h.}
        \label{fig:period_scan_colorbar}
        \end{figure}

        To ensure we do not miss the global minimum in $\chi^2$, we initiated the inversion over a grid of initial spin parameters centered around the values estimated in the previous works. The algorithm then converged to a local minimum in $\chi^2$. The most critical parameters are the two periods; their scan is shown in Fig.~\ref{fig:period_scan}. For each pair of initial $P_\phi$, $P_\psi$, we optimized all eight spin parameters, the shape, and Hapke's scattering parameters. There are several local minima with significantly different periods but having almost the same values of root mean square (RMS) residuals. We selected the 20 best local minima and ran a higher resolution inversion model with more iterations to further improve the fit. This resulted in eight models (listed in Table~\ref{tab:models}) whose periods are marked as dotted lines in Fig.~\ref{fig:period_scan}, and the lower residuals are marked as red asterisks. The plots show a clear pattern of local minima that are separated by $\Delta P_\phi = 0.0106\,$h and $\Delta P_\psi = 0.969\,$h. Figure~\ref{fig:period_scan_colorbar} shows the same data, but now the RMS values are color-coded and the correlation between periods is seen. The three groups of the best local minima are denoted A, B, and C. Their separation is such that over 8 years that separate the apparitions 2012--2013 and 2020--2021, there are $n$ cycles of $P_\phi$ and $m$ cycles of $P_\psi$, where $n \approx 2560$ and $m \approx 270$. All acceptable period pairs follow the relation $(P_{\phi}^{-1} - P_{\psi}^{-1})^{-1} = 30.5658$\,h, which is the main period of Apophis's light curve \citep{Pra.ea:14}. Solution B corresponds to $n+1$ and $m+1$ cycles, while solution C corresponds to $n-1$ and $m-1$ cycles. Because the photometric data can be fitted about equally well with all three combinations of periods, we cannot uniquely determine the number of cycles that elapsed from 2013 to 2020.

\begin{figure*}
        \includegraphics[width=\textwidth]{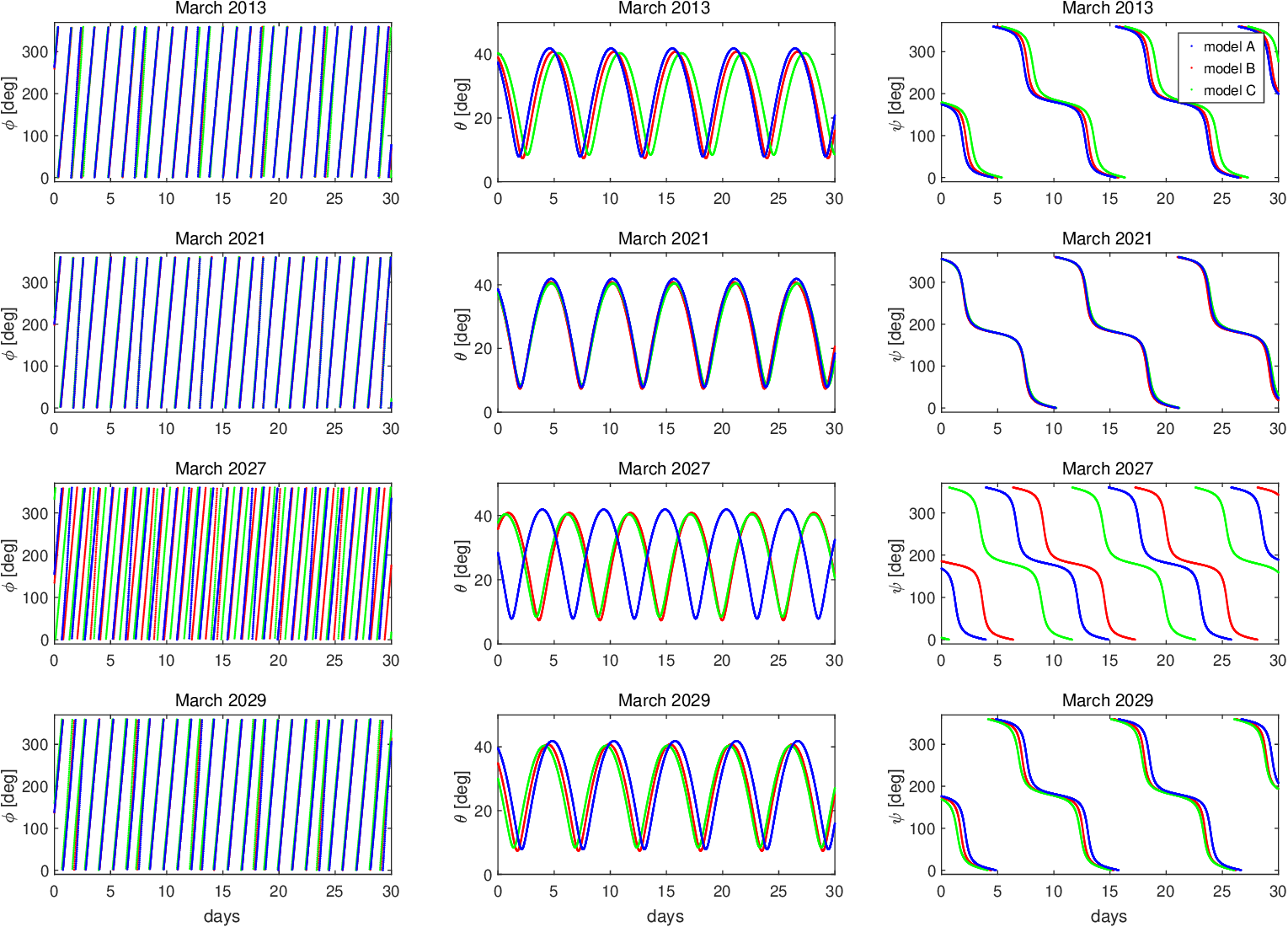}    
        \caption{Evolution of Euler angles for models A (blue), B (red), and C (green). The plots show the evolution of Euler angles $\phi$, $\theta$, and $\psi$ within a month in different years. All three angles are similar for all three models in the years 2013, 2021, and 2029 because they are separated by 8 years. The Euler angles, and thus the orientation of the models, differ for epochs between these years (e.g., March 2027).}
        \label{fig:Euler_angles}
        \end{figure*}

        An interesting feature of this degeneracy is that all three models predict a similar orientation of Apophis in 2029, because the separation between 2013 and 2021 is eight years, which is also the separation between 2021 and 2029. As shown in Fig.~\ref{fig:Euler_angles}, the Euler angles that define the orientation of the body in space are similar for the three models in 2013 and 2021, which is not surprising because the models were derived from observations in these two apparitions. The orientation of these models between and outside the two apparitions is different, which means that we cannot determine it uniquely. However, after another interval of 8 years, the different solutions again give about the same orientation of Apophis. This means that we can predict, at least to some level of precision, the orientation of Apophis during its 2029 encounter. On the other hand, this also means that photometric observations from 2029 cannot effectively distinguish between models A, B, and C. The difference between Euler angles for the three models in April 2029 is $\sim 20^\circ$ in $\phi$, $\sim 5^\circ$ in $\theta$, and $\sim 5^\circ$ in $\psi$. However, as we show in Sect.~\ref{sec:encounter}, the post-encounter spin state is very sensitive to the exact orientation of Apophis, so this accuracy is not sufficient to predict it precisely. 

\begin{figure}[t]
        \includegraphics[width=\columnwidth, trim=1.5cm 6cm 1cm 0.5cm, clip]{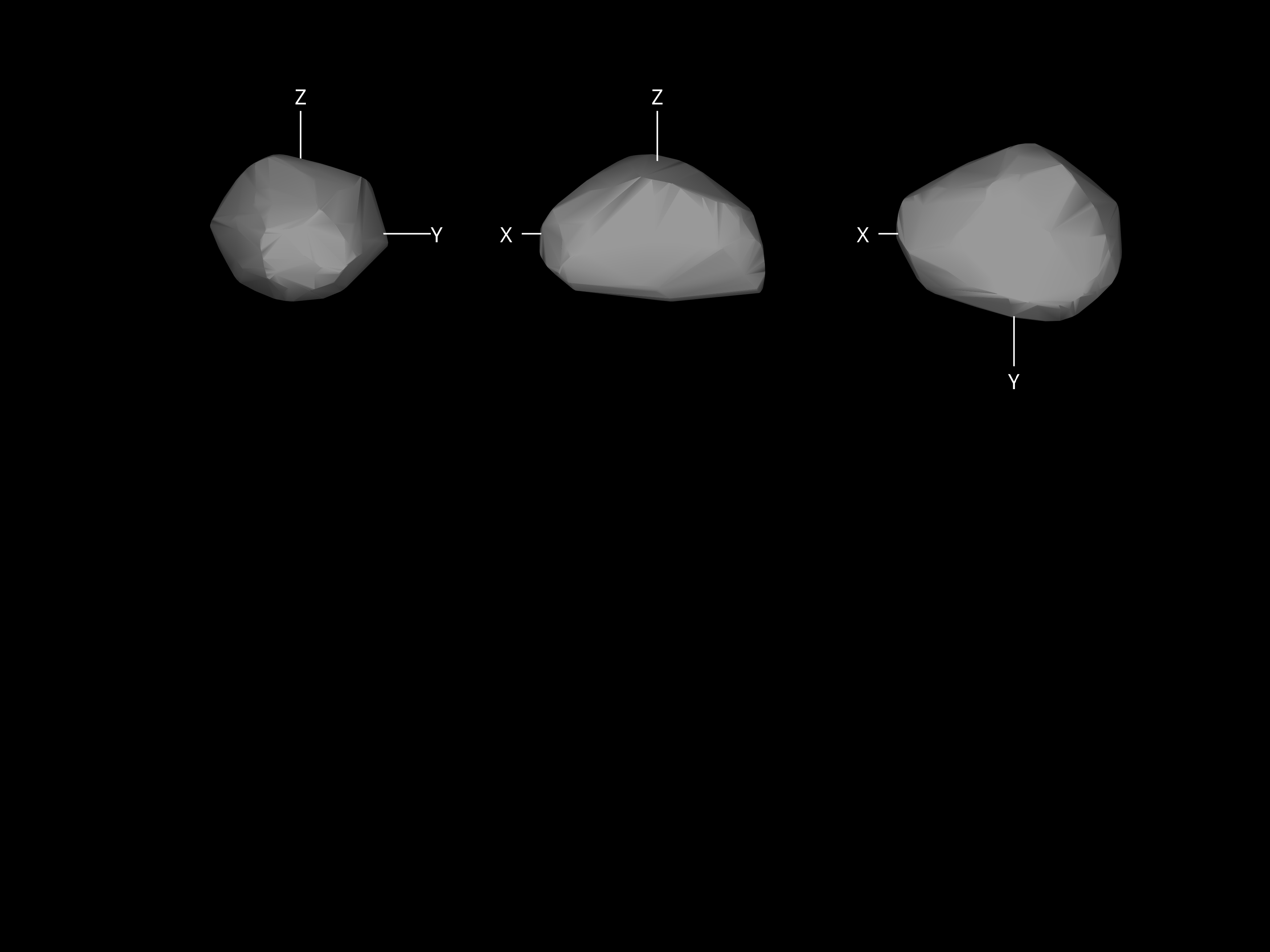} \\   
        \includegraphics[width=\columnwidth, trim=1.5cm 6cm 1cm 0.5cm, clip]{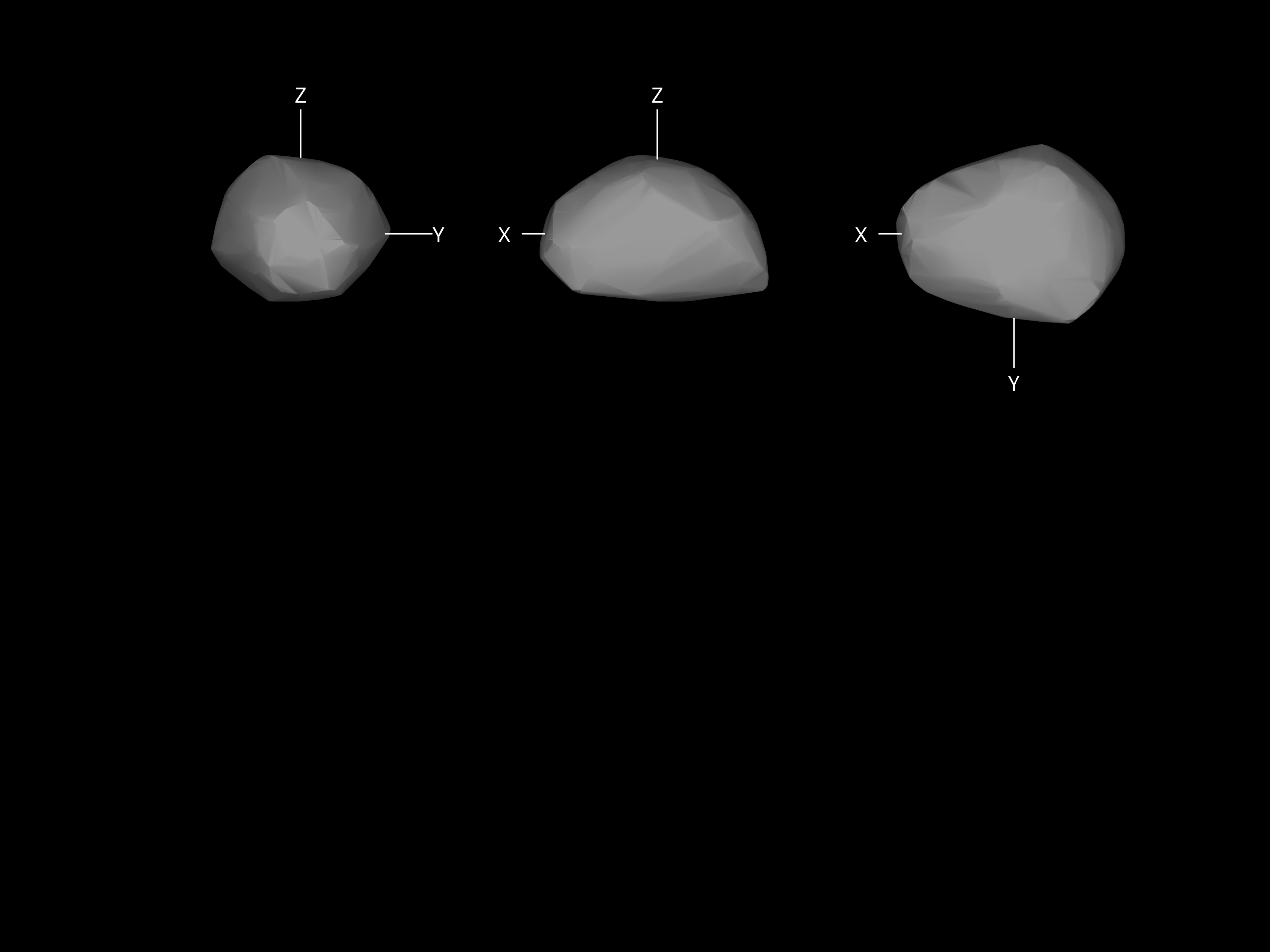}
        \caption{Best models A ({\it top\/}) and A* ({\it bottom\/}) with parameters listed in Table~\ref{tab:best_models}. Model A was constructed from a subset of accurately calibrated photometry from the Danish telescope (La Silla) and TESS observations, while for model B, we used all available photometry.} 
        \label{fig:shape}
        \end{figure}

        The shape model corresponding to solution A is shown in Fig.~\ref{fig:shape}. It is formally the best solution with the lowest RMS residuals. In the same figure, we also show the shape model (called A*) constructed from the full set of photometric data. The visual difference between shape A and A* is about the same as between models A and B or C. Shape models A, A$'$, and A$''$ are practically the same. Figures \ref{fig:lc_fit_2013} and \ref{fig:lc_fit_2021} show how the model fits the photometric data from 2012--2013 and 2020--2021, respectively. The complete spin parameters are listed in Table~\ref{tab:best_models}. The shapes are similar, with minor differences that demonstrate the effect of different datasets on the shape details. We remind the reader that the shape model is convex and that the real shape of Apophis can contain concavities that cannot be reconstructed from the disk-integrated photometry. A bifurcated shape was suggested for Apophis by \cite{Bro.ea:18} based on the bifurcation seen in radar delay-Doppler images taken in 2013. Also, the spin parameters of models A and A* are similar; their differences are order-of-magnitude estimates of the uncertainty caused by the differences between the input datasets. However, both models represent only one local minimum in period parameter space. If we take the spread of periods for models A, $\mathrm{A'}$, and $\mathrm{A''}$ as the estimate of the real periods uncertainty, then the parameters for the best-fit model are $P_\phi = 27.374 \pm 0.001$\,h and $P_\psi = 262.2 \pm 0.1$\,h. The uncertainty intervals are more than one order of magnitude smaller than those of \cite{Pra.ea:14} (due to the much larger time span of our dataset), and they overlap. On the contrary, values reported by \cite{Lee.ea:22} have an order of magnitude smaller formal uncertainties, and they are outside our intervals for both periods. \cite{Lee.ea:22} estimated the uncertainties from $\chi^2$ confidence intervals, which likely correspond to formal errors inside local minima, not to realistic uncertainties over a wider parameter space.

        The error in the direction of the angular momentum vector can be estimated from the standard deviation of red points in Fig.~\ref{fig:pole_nom}, which is $10^\circ$ in the ecliptic latitude $\lambda$ and $0.7^\circ$ in the ecliptic latitude $\beta$. The uncertainty of $\lambda$ seems large, but this is just the effect of spherical coordinates and the $\beta$ value being close to $-90^\circ$. After correcting by $\cos\beta$, the uncertainty in the longitudinal direction becomes $0.6^\circ$, which means that the direction of the angular momentum vector is known with a precision of about $1^\circ$. However, this formal uncertainty does not take into account the effect of regular precession caused by the solar gravitational torque (Sec.~\ref{spin_nominal}). As is apparent from the difference between $\beta$ values for models A and A* in Table~\ref{tab:best_models}, the real uncertainty of the direction of $\vec{L}$ is at least a few degrees.

        Apart from the uncertainty of the periods, there is also a considerable uncertainty of the moments of inertia $I_1$, $I_2$. There is some spread in these values in Table~\ref{tab:models}, but the realistic uncertainty is much larger. Periods can be determined precisely (although not uniquely) because they affect the phase of the light curve, which can be determined precisely from observations. On the contrary, moments of inertia primarily affect the shape of the light curve, which is not fully captured by the data, especially at the beginning and the end of the observing window (Fig.~\ref{fig:lc_fit_2021}), where the light curve sampling is sparse (as the observability of Apophis with ground-based telescopes was limited near the beginning and end of the apparitions). Thus, different combinations of moments of inertia produce slightly different model light curves that still agree with the data. We tested wide intervals of $I_1$, $I_2$, and by assessing how well the synthetic data agreed with the observed data, we roughly estimated the range of possible values as 0.45--0.65 for $I_1$ and 0.965--0.985 for $I_2$, with negative correlation between the parameters. 
        
        Although the parameter $I_1$ is only poorly constrained directly by the model, it is constrained indirectly by the requirement that the kinematic parameters $I_1$, $I_2$ are consistent with principal moments of inertia $I_{1,2}^\mathrm{sh}$ computed ex post from the reconstructed 3D shape model assuming uniform density. Values of $I_{1}^\mathrm{sh}$ for models A and A* are given in Table~\ref{tab:best_models}. They are around 0.6, which is consistent with the range of values from the fitting. Changing $I_{1}^\mathrm{sh}$ to 0.5, for example, would require stretching the shape along the longest axis by about 15\,\%, which would affect the lightcurves significantly. On the other hand, $I_{2}^\mathrm{sh}$ of 0.91 is far outside the uncertainty interval of $I_1$, but bringing it to 0.965 requires less dramatic change of the middle principal axis of about 8\,\%. However, even this shape deformation would clearly deteriorate the light curve fit. Because the 3D shape is not known during the optimization, the shape-dependent parameters $I_{1,2}^\mathrm{sh}$ must be computed after the optimization, serving as a consistency check. Perfect match between $I_{1,2}$ and $I_{1,2}^\mathrm{sh}$ cannot be expected because the convex model inevitably differs in moments of inertia from the real, yet unknown, shape of Apophis. 
        
        \begin{table}
            \caption{Best models.}
            \label{tab:models}
            \centering
            \begin{tabular}{lcccc}
            \hline\hline
            model   & $P_\phi$       & $P_\psi$         & $I_1$     & $I_2$     \\
            \hline
            A                   & 27.3743    & 262.17     & 0.601  & 0.9729  \\ 
            $\mathrm{A'}$       & 27.3737    & 262.12     & 0.585  & 0.9738  \\  
            $\mathrm{A''}$      & 27.3734    & 262.08     & 0.598  & 0.9730  \\[1mm] 
            B                   & 27.3633    & 261.16     & 0.591  & 0.9742  \\ 
            $\mathrm{B'}$       & 27.3637    & 261.20     & 0.588  & 0.9744  \\ 
            $\mathrm{B''}$      & 27.3629    & 261.13     & 0.595  & 0.9733  \\[1mm]
            C                   & 27.3843    & 263.09     & 0.628  & 0.9705  \\  
            $\mathrm{C'}$       & 27.3851    & 263.16     & 0.605  & 0.9722  \\
            \hline        
            \end{tabular}
            \tablefoot{The table lists parameters of the eight best-fit models shown in Fig.~\ref{fig:period_scan}.}
        \end{table}

        \begin{table}
            \caption{Nominal models.}
            \label{tab:best_models}
            \centering
            \begin{tabular}{lcc}
            \hline\hline
            parameter   & model A       & model A* \\
            \hline
            $\lambda$ [deg]  & 246.8       & 248.5 \\
            $\beta$ [deg]    & $-87.2$      & $-83.2$ \\
            $\phi_0$ [deg]   & 133         & 154 \\
            $\theta_0$ [deg] & 19.9        & 27.6 \\
            $\psi_0$ [deg]   & 21          & 193 \\
            $I_1$       & 0.607         & 0.586 \\
            $I_2$       & 0.9724        & 0.9735 \\
            $P_\phi$ [h]   & 27.37430     & 27.37412 \\
            $P_\psi$ [h]   & 262.168       & 262.152\\
            JD$_0$      & 2456300.555   & 2456284.676 \\
            $I_1^\mathrm{sh}$  & 0.605         & 0.612 \\
            $I_2^\mathrm{sh}$  & 0.9145        & 0.9166 \\
            \hline        
            \end{tabular}
            \tablefoot{The table lists parameters of the best models A and A*. Initial Euler angles $\phi_0, \theta_0, \psi_0$ cannot be directly compared as they correspond to different initial times JD$_0$. Parameters $I_{1,2}^\mathrm{sh}$ are normalized principal moments of inertia computed from the shape models (Fig.~\ref{fig:shape}) assuming uniform density.}
        \end{table}
    
    \subsection{Light scattering model}

        To properly model how the brightness of Apophis changes with rotation and changing illumination and viewing geometry, we used Hapke's scattering model \citep{Hap:12} with parameters $w, g, h, B_0, \thetabar$. Because the observations did not cover phase angles below $\sim 20^\circ$, the parameters $h$ and $B_0$ describing the width and amplitude of the opposition effect were fixed at the values for an average S-type asteroid, namely $h = 0.08$, $B_0 = 1.6$ \citep{Hel.Vev:89, Li.ea:15}.\footnote{In practice, our code works with the parameter $S(0) = B_0 w (1-g) / (1 + g)^2$ defined by \cite{Hap:86}, which is for $w = 0.23$, $g = -0.27$ equal to $S(0) = 0.88$, which was the value that we fixed.} Although the observations cover phase angles up to $100^\circ$, the disk-integrated photometry without albedo information does not allow the parameters of Hapke to be constrained. The only parameter that could be reasonably constrained was macroscopic roughness $\thetabar$. The values of $\thetabar$ between 30 and 35 degrees provided the best fit. Parameters $w$ (single-scattering albedo) and $g$ (asymmetry factor) were constrained only poorly to intervals 0.25--0.45 and $-0.3$ to $-0.2$, respectively. The best-fit set of parameters is: $w = 0.42$, $h = 0.08$, $B_0 = 0.94$, $g = -0.25$, and $\thetabar = 33^\circ$. The geometric albedo $p$ is uniquely defined by Hapke's parameters. The values above yield $p = 0.24$, which agrees with the value $0.35 \pm 0.10$ based on the radar shape model \citep{Bro.ea:18} or the range 0.24--0.33 derived from thermal infrared observations \citep{Lic.ea:16}. However, due to the wide range of acceptable Hapke parameters, the range of albedo values is also large, spanning between 0.16 and 0.28.

\section{Apophis's change in spin state during its close approach to Earth in April 2029}
\label{sec:encounter}

\begin{figure}
 \includegraphics[width=\columnwidth]{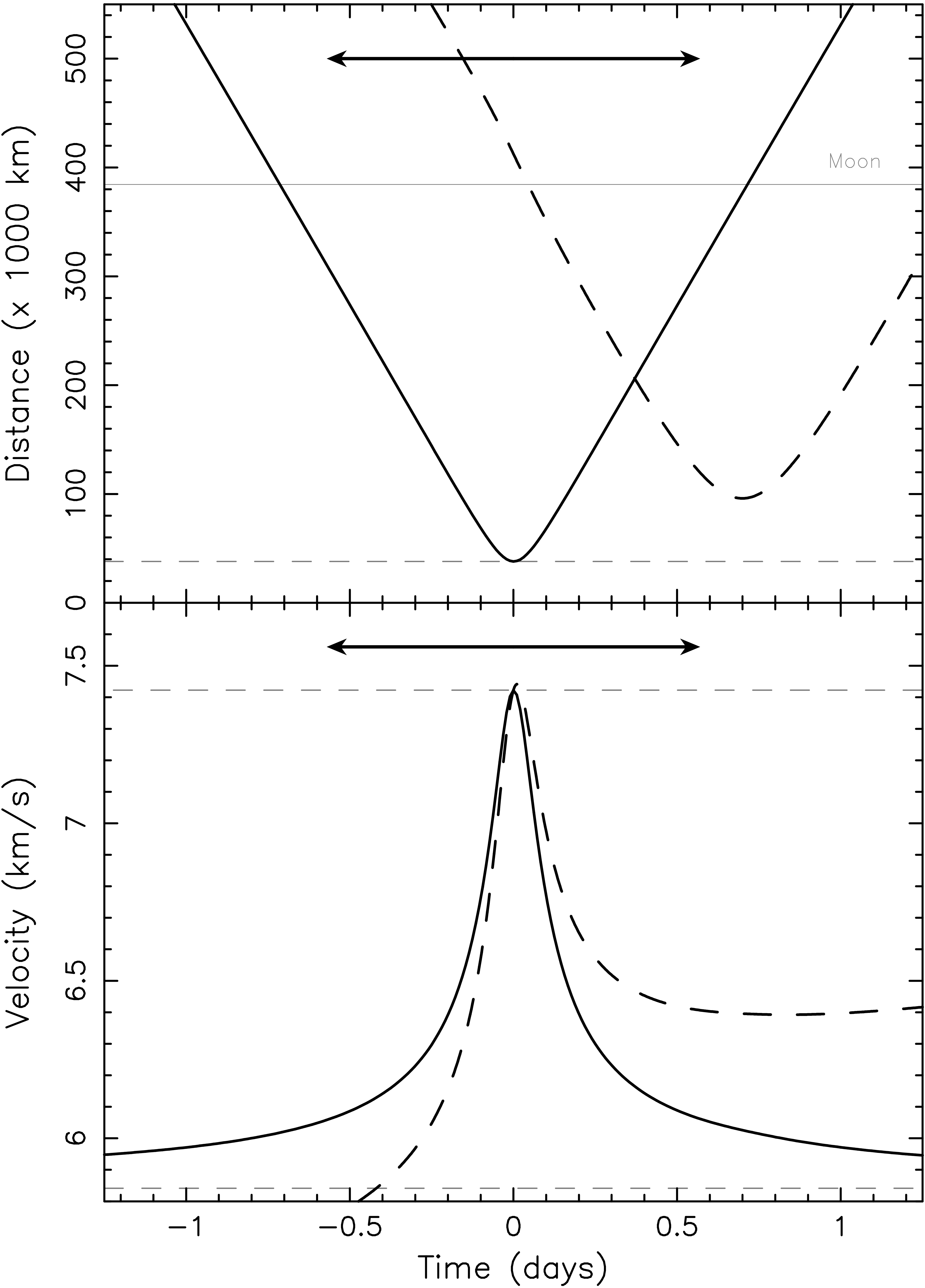}
 \caption{Parameters of the Apophis close encounter with Earth on April~13, 2029. The top panel shows the geocentric (solid line) and selenocentric (dashed line) distance, and the bottom panel shows the relative geocentric and selenocentric velocity. The origin of the abscissa is at the nominal Earth closest approach epoch $62239.90709$ (MJD). The horizontal dashed line at the top panel indicates the closest approach distance of $38012$~km, the horizontal dashed lines at the bottom panel show the relative geocentric velocities at the closest approach, $7.42$~km~s$^{-1}$, and the asymptotic value, $5.84$~km~s$^{-1}$. The closest approach to the Moon is at $95959$~km. The black arrow shows the Apophis $P_\phi$ timescale for reference.}
 \label{fig:enc2029}
\end{figure}

The unusually close approach of the few-hundred-meter sized asteroid Apophis to Earth in April 2029 presents a rather rare event in asteroid science (Fig.~\ref{fig:enc2029}). Not only will it continue to serve as a training case for planetary defense activities 
\citep{2022PSJ.....3..123R}, but sciencewise it will offer a unique possibility to study the physical parameters of this body by the confrontation of various models with observations. While the closest distance of $\sim 6$ Earth radii is still large enough such that the global shape will not significantly change  \citep{2019Icar..328...93D,2021Icar..36514493H}, various studies considered local-scale effects as an alternative way to probe Apophis interior. These include surface mobilization through tidally driven seismic shaking during the encounter and possible surface slope evolution on a longer scale \citep{Yu.ea:14,2019Icar..328...93D,2024PSJ.....5..251B}, or fine structural changes in specific surface regions \citep{2021Icar..36514493H,2023MNRAS.520.3405K}. A necessary pre-requisite of these subtle-phenomena studies consists in the knowledge of Apophis orbital and rotational dynamics before, through, and immediately after the encounter
\citep[the latter being also a crucial information for the OSIRIS-APEX operations; e.g.,][]{2023PSJ.....4..198D}. The orbital part is presently known with sufficient accuracy \citep{2022LPICo2681.2023B,2024LPICo3006.2046F}, but the precision of the rotational part is far poorer. 
While radar observations allow for improvements in the rotational state of Apophis \citep[e.g.,][]{2022LPICo2681.2023B}, their main strength lies in orbit and shape characterization. Spin state models based on analysis of the photometric data still provide state-of-art information. In the previous sections, we argued that our results here improve on the quality of the model presented by \citet{Lee.ea:22}. As a result, we take the opportunity and present a full-fledged model of Apophis spin evolution through the 2029 Earth encounter.

This goal has been considered in a number of papers ever since the Apophis close approach to Earth in 2029, and its slow rotation has been determined. An early effort in this respect is due to \citet{2005Icar..178..281S}. More accurate analyses followed the discovery of Apophis's tumbling state by \citet{Pra.ea:14}. However, the uncertainty of their solution based on the Apophis photometry from the single apparition 2012--2013 expands too much by 2029, such that only weak constraints of the Apophis spin state change during and after the 2029 encounter could be achieved \citep[e.g.,][]{2023Icar..39015324B}. Some authors, therefore, conducted analyses using statistical sampling of plausible pre-encounter rotation state only \citep{2018A&A...617A..74S,2019Icar..328...93D,2024SoSyR..58..208L}. Luckily, things improve with our solution of the Apophis spin state. Its nominal pre-encounter phase uncertainty shrunk considerably, such that the post-encounter state is formally much better constrained \citep[within the model assumptions
similar to][]{2021Icar..36514493H}. Results from this nominal model are presented in Sec.~\ref{spin_nominal}.

However, we identify a possible caveat in the absence of the radiative torques, the Yarkovsky-O'Keefe-Radzievskii-Paddack (YORP) effect in particular \citep[e.g.,][]{Rub:00,
Bot.ea:06,Vok.ea:15}, in the Apophis spin state modeling. The present photometric data from the two apparitions do not allow for modeling of the possible YORP effect in the asteroid. Its dynamical consequences may still be represented by a parameter change of the nominal model. However, the 2029 encounter acts in this respect as a third apparition when the solution of the nominal model and that with the radiative torques included diverge. In Sec.~\ref{spin_YORP}, we estimate the extent of the spin state uncertainty due to the missing YORP torques and argue that the pre-encounter ground-based observations in 2027 or 2028 could significantly improve the solution
\citep[or adding a further rationale to Apophis space-based observational efforts before the 2029-Earth encounter such as FLARE; e.g.,][]{2024LPICo3006.2060B}.

Our model has already been validated and used for the analysis of the rotation state of a small asteroid 2012~TC4 by \citet{Lee.ea:21}. For that reason, we refer to this work, in particular its Appendix, for detailed information, while in the next Section, we briefly recall our approach.

\subsection{Apophis spin state propagation} \label{prop}
The kinematical part of the spin evolution describes 
orientation of the asteroid in the inertial frame. We assume the asteroid is a rigid body (valid even through the 2029 encounter), allowing us to define unambiguously a proper body-fixed frame. The easiest choice has (i) the origin in the asteroid's center-of-mass, and (ii) the axes coinciding with the principal axes of the inertia tensor $\mathbf{I}$ (therefore $\mathbf{I}={\rm diag}(A,B,C)$, with $A\leq B\leq C$). In fact, the photometric data fitting procedure used the free-top model with these assumptions built in. The transformation between the inertial and body-fixed frames is conventionally parametrized by a set of Euler angles, most often the 3-1-3 sequence of the precession angle $\phi$, the nutation angle $\theta$, and the angle of proper rotation $\psi$. However, instead of the three Euler angles $(\phi,\theta,\psi)$ we use here the Rodrigues-Hamilton parameters $\mbox{\boldmath$\lambda$}=(\lambda_0,\lambda_1,\lambda_2,\lambda_3)$ \citep[e.g.,][]{Whi:1917}, also used by some of the previous publications dealing with Apophis spin state \citep[e.g.,][]{2021Icar..36514493H,2024SoSyR..58..208L}. Their relation to the Euler angles is given by: (i) $\lambda_0+\imath \lambda_3 = \cos\frac{\theta}{2}\,\exp\left[\frac{\imath}{2} \left(\psi+\phi\right)\right]$, and (ii) $\lambda_2+\imath \lambda_1= \imath \sin\frac{\theta}{2}\,\exp\left[\frac{\imath}{2} \left(\psi-\phi\right)\right]$ ($\imath=\sqrt{-1}$ is a complex unit). One can easily verify a constraint: $\lambda_0^2+
\lambda_1^2+\lambda_2^2+\lambda_3^2=1$ (in our numerical runs satisfied with a $\leq 10^{-13}$ accuracy). 
The rotation matrix $\mathbf{A}$ needed for the vector transformation from the inertial frame to the body-fixed frame is a simple quadratic form of $\boldsymbol{\lambda}$, namely 
\begin{equation}
 \begin{split}
 & \mathbf{A} = \\
 & \left(
 \begin{array}{lll}
 \lambda_0^2+\lambda_1^2-\lambda_2^2-\lambda_3^2 &  2\left(\lambda_0\lambda_3+
  \lambda_1\lambda_2\right) &  2\left(\lambda_1\lambda_3-\lambda_0\lambda_2\right) \\
 2\left(\lambda_1\lambda_2-\lambda_0\lambda_3\right) &  \lambda_0^2+\lambda_2^2-
  \lambda_1^2-\lambda_3^2 & 2\left(\lambda_0\lambda_1+ \lambda_2\lambda_3\right) \\
 2\left(\lambda_0\lambda_2+ \lambda_1\lambda_3\right) & 2\left(\lambda_2\lambda_3-
  \lambda_0\lambda_1\right) & \lambda_0^2+\lambda_3^2-\lambda_1^2-\lambda_2^2 \\
 \end{array}
 \right) \; . 
 \end{split}
 \label{e1}
\end{equation}
The inverse transformation is represented by a transposed matrix $\mathbf{A}^{\rm T}$. Asteroid's rotation is represented with the angular velocity vector $\boldsymbol{\omega}$, whose components in the body-fixed frame are $(\omega_1,\omega_2,\omega_3)$. Their relation to the time derivatives of the Rodrigues-Hamilton parameters is simply
\begin{equation}
 \frac{\dd\boldsymbol{\lambda}}{\dd t} = \frac{1}{2}\,\mathbf{P}\cdot 
  \boldsymbol{\lambda} \; , \label{e2}
\end{equation}
where
\begin{equation}
 \mathbf{P} = \left(
 \begin{array}{cccc}
  0 &  -\omega_1 &  -\omega_2 &  -\omega_3 \\
   \omega_1 &  0 & \phantom{-}\omega_3 &  -\omega_2 \\
   \omega_2 &  -\omega_3 &  0 &  \phantom{-}\omega_1 \\
   \omega_3 & \phantom{-}\omega_2 &  -\omega_1 &  0 \\
 \end{array}
 \right) \; . \label{e3}
\end{equation}

The dynamical part of the problem expresses the Newton's principle that a change of the rotational (intrinsic) angular momentum $\vec{L} = \mathbf{I}\cdot \boldsymbol{\omega}$ is given by the applied torque $\vec{M}$. Tradition has it to state this rule in the body-fixed frame, where $\mathbf{I}$ is constant and even diagonal in our choice of axes, such that
\begin{equation}
 \frac{\dd\vec{L}}{\dd t} + \mbox{\boldmath$\omega$}\times \vec{L} = \vec{M} \; . \label{e4}
\end{equation}
Equations (\ref{e2}) and (\ref{e4}) define the problem of asteroid's rotation in our set of seven parameters $(\mbox{\boldmath$\lambda$},\mbox{\boldmath$\omega$})$. Once the torques $\vec{M}$ are specified, we numerically integrate this system of differential equations with the initial data determined at  1~January 2021 provided by our solution in Sec.~\ref{sec:model}. We use Burlish-Stoer integration scheme with tightly controlled accuracy.
\begin{figure*}
 \begin{center}
  \includegraphics[width=0.9\textwidth]{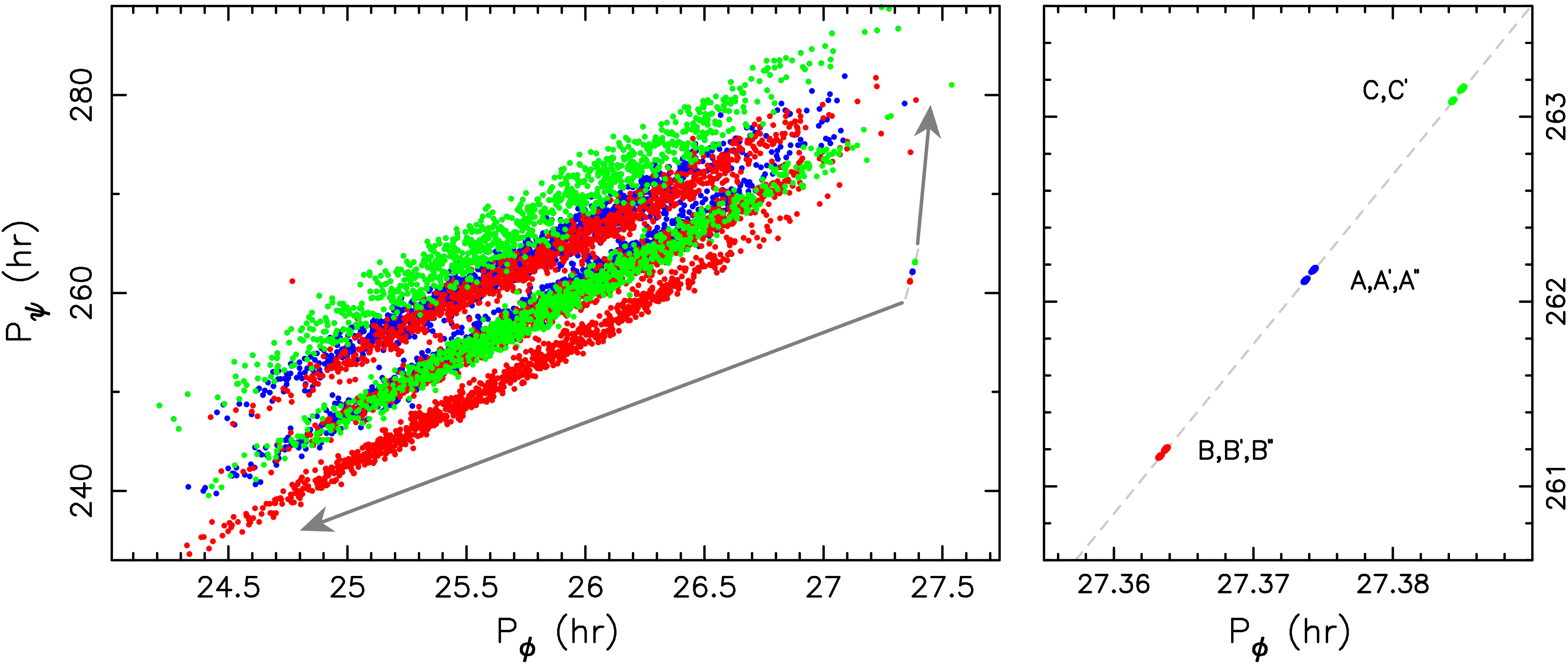}    
 \end{center}
 \caption{Rotation $P_\psi$ and precession $P_\phi$ periods of Apophis before and after the close encounter with Earth in April 2029 calculated from the purely gravitational model of the spin state change. Individual symbols come from 1000 nominal model simulations, with gravitational torques included, using statistically equivalent initial conditions on January~1, 2021. The three colors identify solutions starting from the three statistically equivalent initial solutions (compare with Fig.~\ref{fig:period_scan}): red for the B-group, blue for the A-group, and green for the C-group. The right panel shows the current accuracy of the Apophis solution from the combination of the 2012--2013 and 2020--2021 observations. Those data are also reproduced in the left panel, but with a larger abscissa and ordinate scale, they are resolved only as single dots. The post-encounter $P_\phi$ and $P_\psi$ values, spread over a large region, are shown on the left panel only. As expected, the uncertainty interval in both quantities expands by a factor larger than $10^3$. The post-encounter median $P_\phi$ value decreases to $\simeq 25.75$~hr, while the median $P_\psi$ value remains about the same.}
 \label{fig:periods_nom}
\end{figure*}

The energy of rotational motion about the center is given by $E=\frac{1}{2}\,\vec{L}\cdot\boldsymbol{\omega}$. In a classical problem of a free top (i.e., $\vec{M}=0$), both $E$ and $\vec{L}$ in the inertial frame are conserved. In the body-fixed frame, only $L=|\vec{L}|$ is constant. Nevertheless, the conservation of $E$ and $L$ uniquely determines the wobbling trajectory of $\vec{L}$ in the body-fixed frame \citep[e.g.,][]{Lan.Lif:69,Dep.Eli:93}. There are two options for this motion: (i) short-axis mode (SAM), when $\vec{L}$ circulates about $+z$ or $-z$ body axis, or (ii) long-axis (LAM), mode $\vec{L}$ circulates about $+x$ or $-x$ body axis. A useful discriminator of the two is yet another conserved and dimensionless quantity in the free-top problem, namely $p=2BE/L^2$: (i) SAM is characterized by $p$ values in between $\beta=B/C$ and $1$, while (ii) LAM is characterized by $p$ values in between $1$ and $\alpha=B/A$. Note that $\Delta=B/p$ plays an important role in the description of the free-top problem using Hamiltonian tools \citep[e.g.,][]{Dep.Eli:93,Bre.ea:11}. The free-top motion of $\vec{L}$ in the body fixed frame is easily integrable using Jacobi elliptic functions. When plugged in the kinematical equations (\ref{e3}), one can also obtain solution for $\boldsymbol{\lambda}$ or the Euler angles $(\phi,\theta,\psi)$ \citep[e.g.,][]{Whi:1917,
Lan.Lif:69}. Those of $\psi$ and $\theta$ are strictly periodic, with a period (SAM mode relevant for Apophis assumed here)
\begin{equation}
 P_\psi = \frac{C}{L}\, \frac{4\beta K\left(k\right)}{\sqrt{\left(1-\beta\right) \left(\alpha-p\right)}}\; , \label{e5}
\end{equation}
where $K(k)$ is a complete elliptic integral of the first kind with the modulus $k$ 
given by
\begin{equation}
 k^2 = \frac{\left(\alpha-1\right)\left(p-\beta\right)}{\left(1-\beta\right)\left(\alpha-p\right)}  \label{e6}
\end{equation}
(the motion of $\theta$ has a periodicity $P_\psi/2$). The motion of the precession angle $\phi$ is not periodic. Nevertheless, a fully analytical solution still exists and it is composed of two parts, the first of which has periodicity $P_\psi$ and a second has another periodicity, generally incommensurable with $P_\psi$ \citep[e.g.,][]{Whi:1917,Lan.Lif:69}. Yet it is both practical and conventional to define an approximate periodicity $P_\phi$ of $\phi$ by numerically computing the number of cycles of $\phi$ over a sufficiently large time interval $T>P_\phi$. In the case of Apophis rotation, we used $T=3\,P_\psi$ (a little more than a month, or some 27 $P_\phi$ cycles). 
It is also useful to note that the minimum $\theta_-$ and maximum $\theta_+$ values of the nutation angle during the rotation period $P_\psi$ are given by $\cos^2\theta_-=(\alpha-p)/(\alpha-\beta)$ and $\cos^2\theta_+=(1-p)/(1-\beta)$ (SAM mode). The mean value of $\cos\theta$ over $P_\psi$ cycle is $\frac{\pi}{2\,K(k)}\sqrt{\frac{\alpha-p}{\alpha-\beta}}$.
When weak torques are applied,  the free-top solution still represents a very useful (osculating) template with all the above-discussed variables such as $E$, $L$, $p$, $P_\psi$, or $P_\phi$ adiabatically changing in time.
\begin{figure}
 \includegraphics[width=\columnwidth]{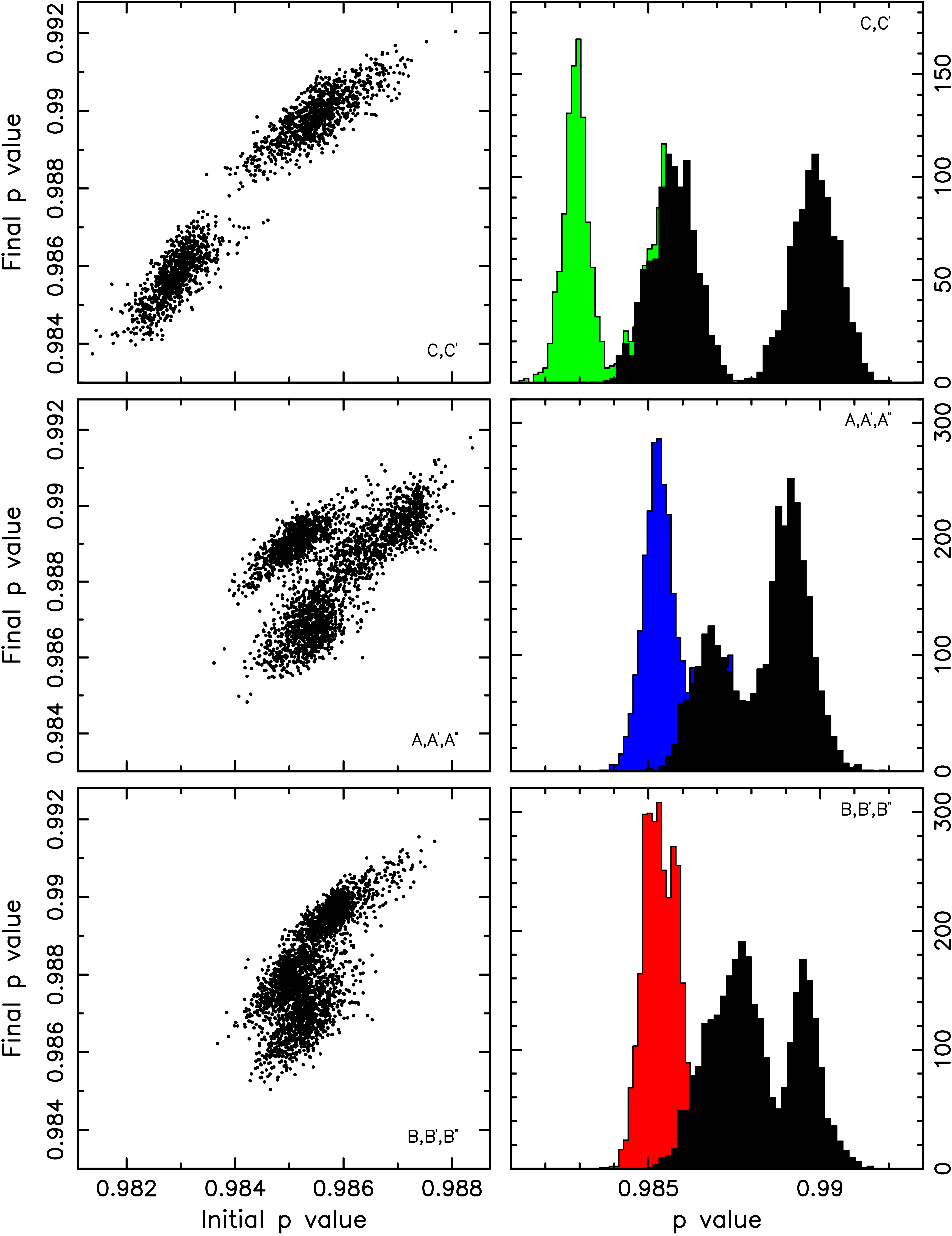}
 \caption{Dimensionless parameter $p=2BE/L^2$ of Apophis's spin. Left panel: Pre-encounter values at the abscissa and the post-encounter values, calculated from the purely gravitational model of the spin state change, at the ordinate. As in the previous two figures, symbols come from 1000 nominal model simulations. Right panel: Distribution of the $p$-values, pre-encounter in green, blue, and red for models C, A, and B, respectively, and post-encounter in black. As all values are smaller than unity, the rotation state remains in the SAM mode. The overall increase of the post-encounter values, however, implies the rotation stepped towards the separatrix of the SAM and LAM modes.}
 \label{fig:pint_nom}
\end{figure}
\begin{figure}
 \includegraphics[width=\columnwidth]{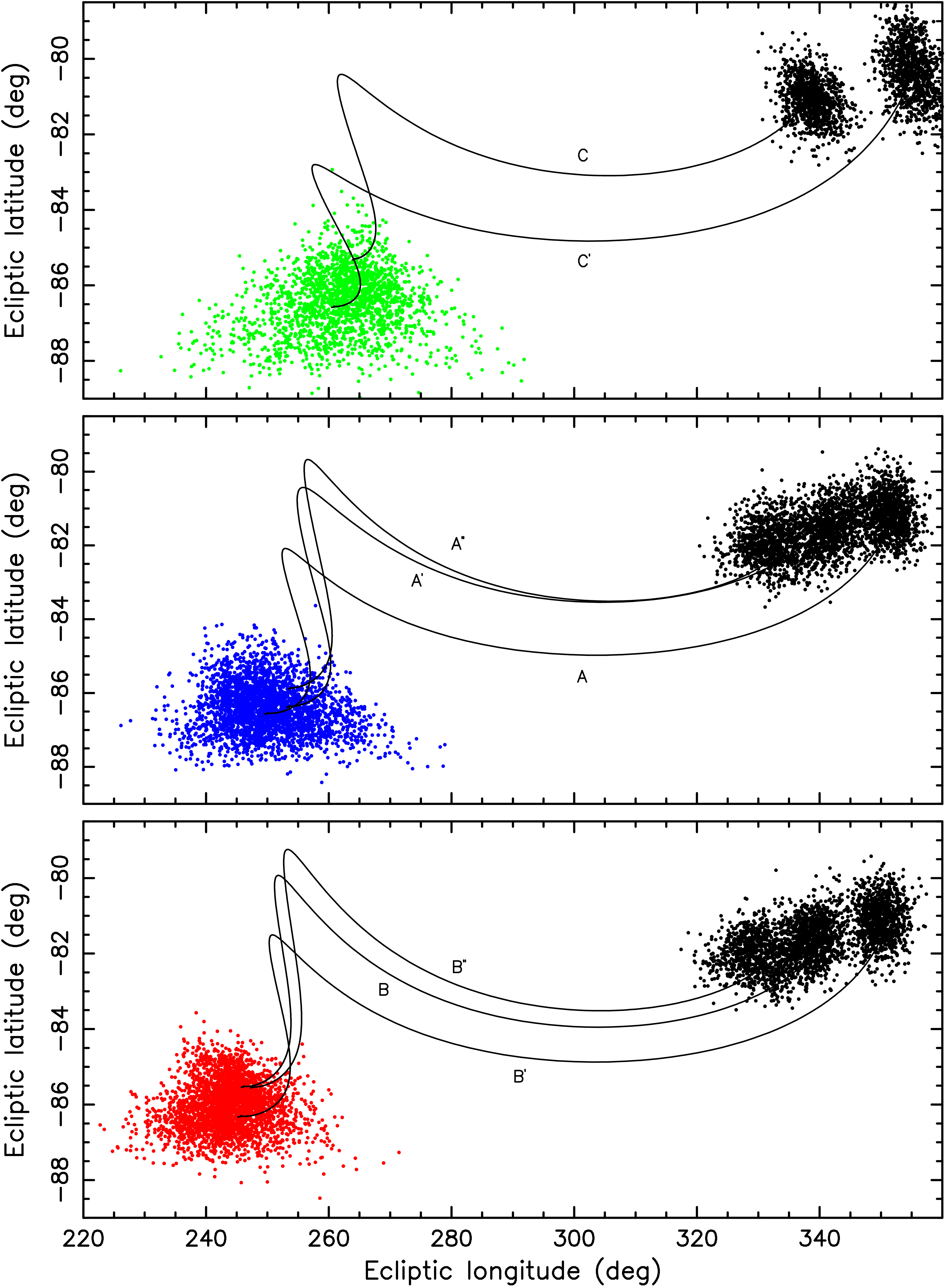}
 \caption{Ecliptic longitude and latitude of the Apophis rotation angular momentum ${\vec{L}}$ before (green, blue, and red for models C, A, and B, respectively) and after (black) the close encounter to Earth in April 2029, calculated from the purely gravitational model of the spin state change.  Individual symbols come from 1000 nominal model simulations, with gravitational torques included, using statistically equivalent initial conditions on 1 January 2021. All of the action, illustrated for our best-fit solution by the black line, occurs within $\simeq \pm 2$~days centered at the Earth encounter. While the median longitude changes by more than $100^\circ$, the true tilt of ${\vec{L}}$ is far smaller because of the near-polar latitude. Its median value is $\simeq 9.8^\circ$. Unlike the periods $P_\phi$ and $P_\psi$ in Fig.~\ref{fig:periods_nom}, the post-encounter uncertainty region in longitude and latitude is not significantly stretched.}
 \label{fig:pole_nom}
\end{figure}

Finally, we discuss the torques used in our analysis. The first class is due to the gravitational tidal fields of the Sun and Earth.%
\footnote{As Figure~\ref{fig:enc2029} shows Apophis will approach the Moon about $16$~hours after the closest encounter with Earth. However, we do not include this body in our simulation at this moment. Not only is the lunar mass nearly two orders of magnitude smaller than that of Earth, but also the closest distance to the Moon will be about $2.5$ times larger than to Earth. As a result, the maximum of the lunar torque on Apophis spin will be $\simeq 1300$ times smaller.}
This constitutes our nominal model, whose results are discussed in Sec.~\ref{spin_nominal}. We assumed a point mass source $M$ specified in the body-fixed frame of the asteroid with a position vector $\vec{R}$ (the non-sphericity of the perturber may be safely neglected even for the Earth encounter in 2029). Using the quadrupole part of the exterior perturber tidal field in the body-fixed frame, we have \citep[e.g.,][]{Fit:70,Tak.ea:13}
\begin{equation}
 \vec{M}_{\rm grav} = \frac{3GM}{R^5}\,\vec{R}\times \left(\mathbf{I}
  \cdot\vec{R} \right)\; . \label{e7}
\end{equation}
We neglected the formally dipole part of the tidal field, which could only occur if the true center-of-mass of the asteroid is slightly displaced from the assumed location \citep[determined by using the assumption of homogeneous density; see, e.g.,][]{Tak.ea:13}. The positions of all bodies, the asteroid, the Sun, and the Earth, are primarily determined using the numerical integration of the orbital problem in the inertial frame. In our case, we numerically integrated planetary orbits, including Earth and Apophis, in the heliocentric system by taking initial data from the NEODyS website\footnote{\url{https://newton.spacedys.com/neodys/}}. We output the necessary positions every $15$~minutes, enough for the purpose of Apophis's rotation dynamics. We also compared our solution with that available at the JPL Horizons system\footnote{\url{http://ssd.jpl.nasa.gov/?horizons}}, and found a very good correspondence with tiny differences, not meaningful for our application.\footnote{The times of the closest approach differed by 0.5\,s; the distances at the closest approach by 1\,km.} The relative position $\vec{R}$ in (\ref{e7}) is determined by (i) the  difference of the corresponding bodies in our orbital solution, and (ii) transformation to the body-fixed
frame. As a result $\vec{M}_{\rm grav} = \vec{M}_{\rm grav}\left(t,
\boldsymbol{\lambda}\right)$. The steep dependence $M_{\rm  grav}\propto R^{-3}$ implies that the Earth effect is non-negligible only during the close encounters with this planet (Fig.~\ref{fig:enc2029}). 

We also extended the nominal model by additionally including radiative torque, in particular, the effect due to sunlight reflection and thermal re-radiation by the asteroid surface. These phenomena are generally termed the YORP effect \citep[e.g.,][]{Rub:00,Bot.ea:06,Vok.ea:15}. 
The justification for this extension is twofold. First, the available detections of the YORP effect for small near-Earth asteroids \citep[see Table~1 in][]{2024A&A...682A..93D} allow us to roughly estimate the perturbation. Consider, for instance, the case of a similarly sized S-type asteroid (25143) Itokawa, for which the available observations reveal an acceleration of the rotation rate $\upsilon\simeq 4\times 10^{-8}$~rad~day$^{-2}$. Itokawa's orbit is slightly exterior to that of Apophis, and the inferred value $\upsilon$ is, in fact, nearly an order of magnitude smaller than predicted from simpler variants of the YORP effect modeling \citep[e.g.,][]{Sch.ea:07,Bre.ea:09}. As a result, the $\upsilon$ value determined for Itokawa may plausibly be a conservative guess of the value expected for Apophis. If true, in $T\simeq 17$~years, from 2012 to 2029, the YORP effect may contribute by $\frac{1}{2}\upsilon T^2\simeq 45^\circ$ in the Apophis rotation phase. Such a value would hugely change the prediction of the spin change during the 2029 encounter (even a few times smaller effect would still be relevant). Note that the tumbling rotation state does not prevent YORP from operating, only it makes its modeling more complicated, calling for numerical analysis \citep[e.g.,][]{Vok.ea:07} or simplified analytical models \citep[e.g.,][]{Cic.Sch:10,Bre.ea:11}. Additionally, it is conceivable that the observed change in the tumbling state of the very small near-Earth asteroid 2012~TC4 is due to the YORP effect \citep{Lee.ea:21}. Secondly, the Yarkovsky effect (an alter ego phenomenon to YORP acting on the orbital dynamics) has been determined for Apophis with a very high accuracy \citep[in fact, it represents the second most precise detection after Bennu]{2022LPICo2681.2007F,2024LPICo3006.2046F}. This again supports YORP as an active component in Apophis rotational dynamics.
\begin{figure*}
 \begin{center}
  \includegraphics[width=0.9\textwidth]{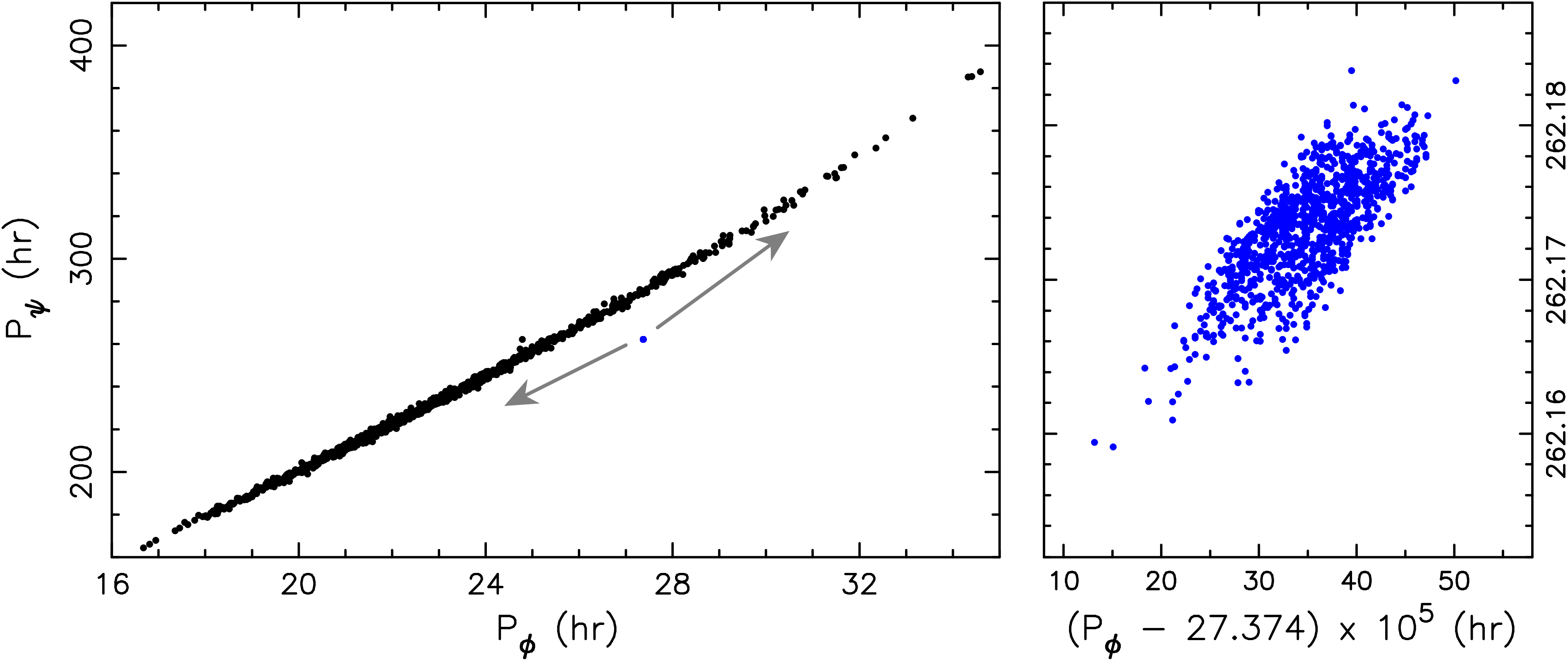}    
 \end{center}
 \caption{Same as in Fig.~\ref{fig:periods_nom} and model A, but now for a set of simulations including the YORP torque (\ref{e8}) with a fudge factor $f=0.4$. Stretching of both $P_\phi$ and $P_\psi$ periods is now larger by a factor of a few.}
 \label{fig:periods_yorp}
\end{figure*}
\begin{figure}
 \includegraphics[width=\columnwidth]{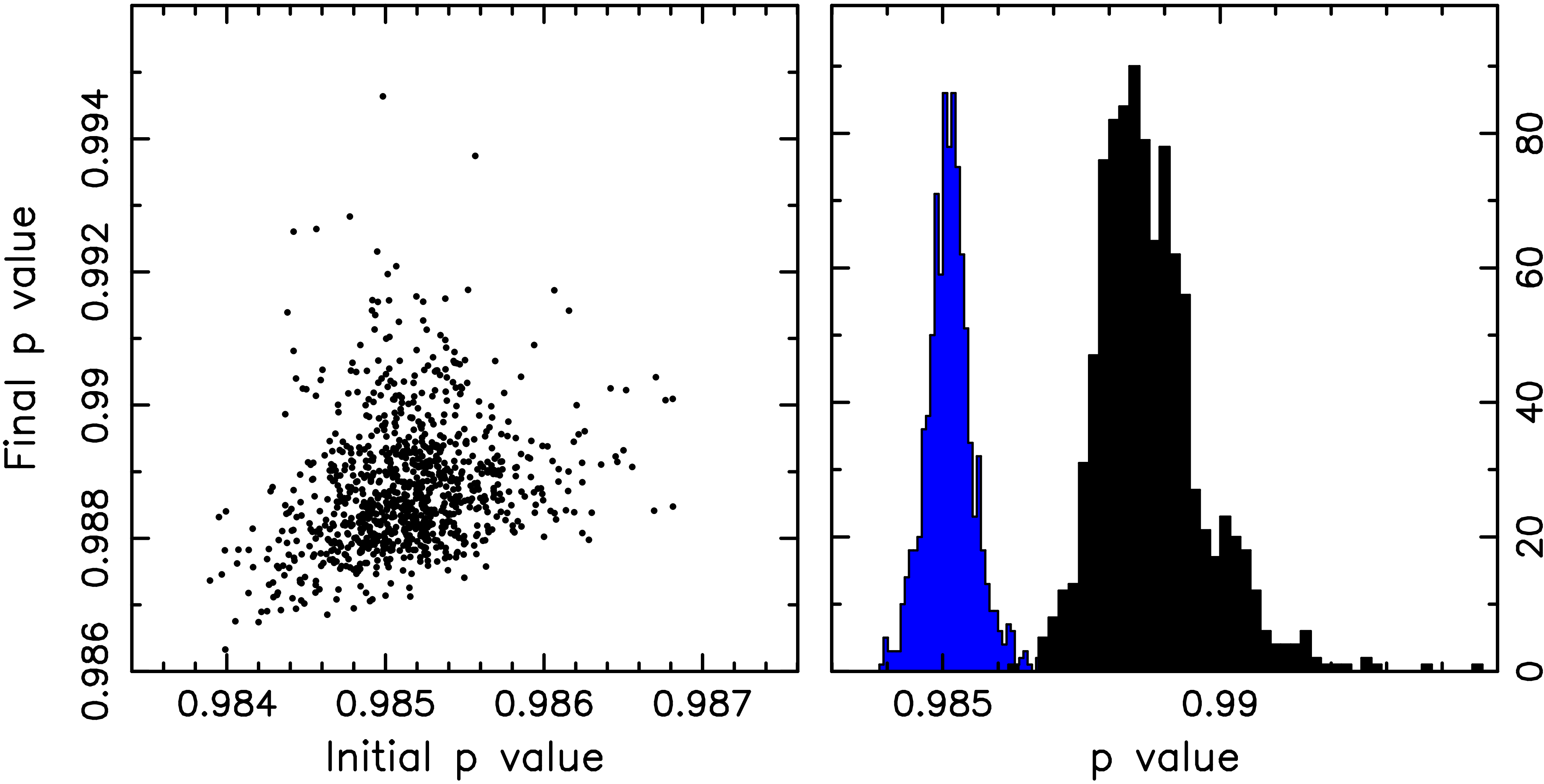}
 \caption{Same as in Fig.~\ref{fig:pint_nom} and model A, but now for a set of simulations including the YORP torque (\ref{e8}) with a fudge factor $f=0.4$. The post-encounter $p$-values now exhibit a tail towards unity, which is a separatrix between SAM and LAM modes.}
 \label{fig:pint_yorp}
\end{figure}
\begin{figure}
 \centering\includegraphics[width=\columnwidth]{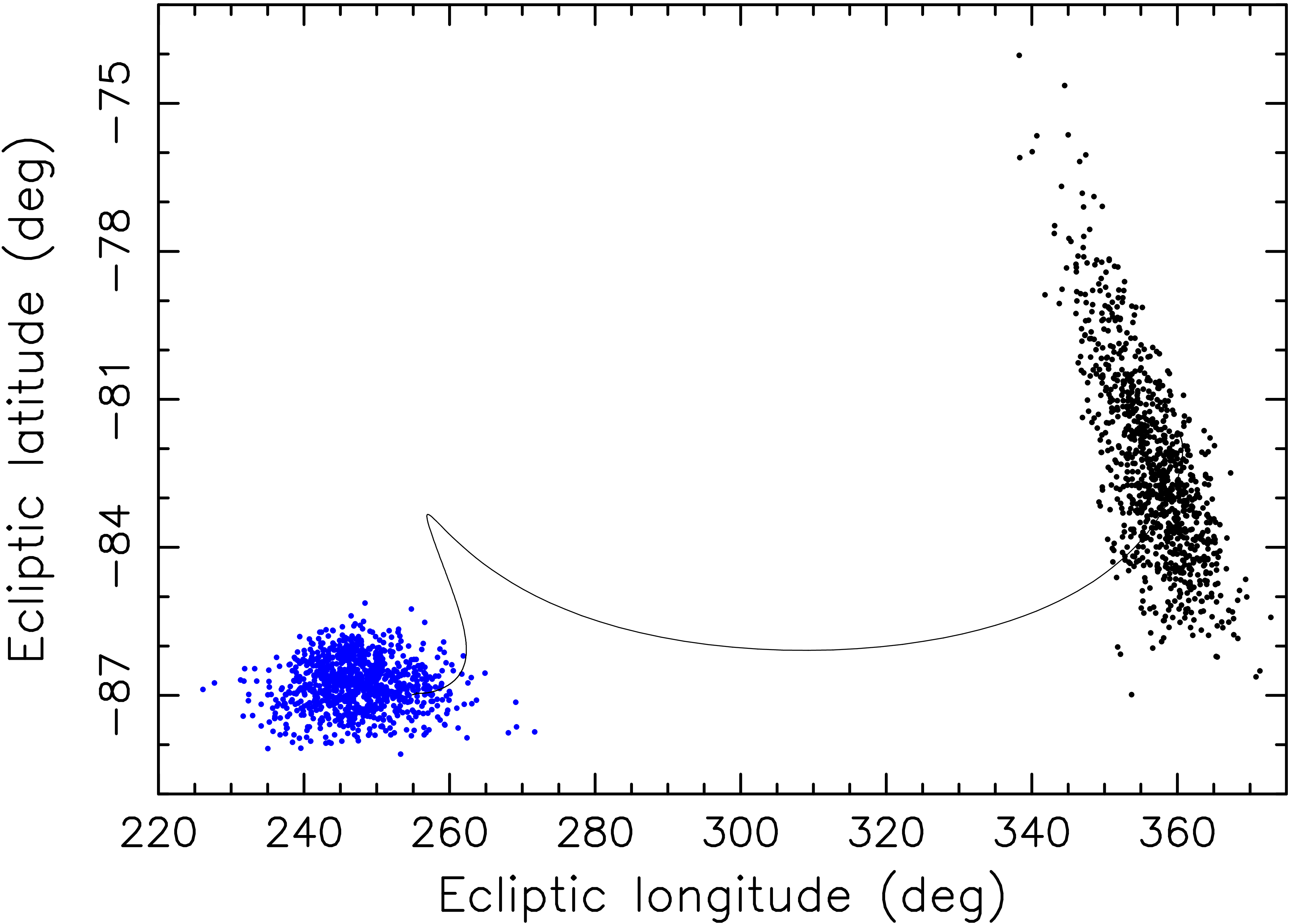}
 \caption{Same as in Fig.~\ref{fig:pole_nom} and model A, but now for a set of simulations including the YORP torque (\ref{e8}) with a fudge factor $f=0.4$. The post-encounter pole latitude values now stretch beyond $-80^\circ$.}
 \label{fig:pole_yorp}
\end{figure}

Because of the tumbling rotation state of Apophis, we resorted to the simplest variant, namely a limit of zero thermal inertia \citep[the effects of finite thermal inertia were studied only for objects rotating about the shortest axis of the inertia tensor so far; e.g.,][]{Cap.Vok:04,Gol.Kru:12}. In this approximation, the radiation torque is given by \citep[e.g.,][]{Rub:00,Vok.Cap:02}
\begin{equation}
 \vec{M}_{\rm YORP}= -\frac{2 F f}{3c}\int H\left(\vec{n}\cdot\vec{n}_0\right)  \left(\vec{n}\cdot\vec{n}_0\right)\,\vec{r}\times \dd\vec{S}\; , \label{e8}
\end{equation}
where $F$ is the solar radiation flux at the location of the asteroid and $c$ is the light velocity. The integral in (\ref{e8}) is performed over the surface of the body characterized by an ensemble of outward-oriented surface elements $\dd\vec{S}=\vec{n}\,\dd S$, $\vec{n}$ is the normal to the surface and $\vec{r}$ is the position of the surface element with respect to the origin of the body-fixed frame. The unit vector of the solar position in the body fixed frame is denoted with $\vec{n}_0$, and $H(x)$ is the Heaviside step function (its presence in the integrand of (\ref{e8}) implies that a non-zero contribution to the radiation torque is provided by surface units for which the Sun is above the local horizon). In fact, our code includes even more complex feature of self-shadowing of surface units, but this is not active in the case of Apophis whose resolved shape is convex. The factor $2/3$ on the right-hand side of Eq.~(\ref{e8}) is due to the assumption of Lambertian reradiation from the surface. The lightcurve inversion obviously allows only a finite accuracy in shape determination of the body, typically a convex polyhedron with little more than a thousand surface facets. The formal integration in Eq.~(\ref{e8}) is therefore represented with a summation over the surface units of the resolved shape model. We used algebra from \citet{Dob:96} to determine all the necessary variables. This also means we assumed a constant density
distribution in the body. Finally, the constant $0\leq f\leq 1$ in (\ref{e8}) is a "fudge factor" we arbitrarily introduce in $\vec{M}_{\rm YORP}$. The expected values smaller than unity should simulate the often observed exaggeration of the YORP torque in the simplest, zero-conductivity limit \citep[see discussion in][]{Vok.ea:15}.

\subsection{Results}
The initial rotation state and shape models of Apophis were constructed by bootstrapping observations from the 2012--2013 and 2020--2021 apparitions. We created 1000 such variants and set the epoch to January~1, 2021. These solutions were propagated using the model described in Sec.~\ref{prop} with the final epoch in September~2029, well beyond the close approach to Earth in March~2029. We proceed in two steps. In Sec.~\ref{spin_nominal} we first include only gravitational torques in the analysis. Next, in Sec.~\ref{spin_YORP}, we additionally demonstrate the potential role of the YORP torque.

\subsubsection{Purely gravitational model}\label{spin_nominal}
Figure~\ref{fig:periods_nom} shows the predicted change in the two fundamental tumbling periods, namely the precession periods $P_\phi$ at the abscissa and the rotation period $P_\psi$ on the ordinate. Similarly to the close encounter effect on the Apophis orbit, namely a huge stretch of the uncertainty interval in the osculating elements, the uncertainty in both periods significantly expands (by more than a factor $10^3$). Still, the present fair constraint on their pre-encounter values makes the post-encounter uncertainty reasonably controlled. We find that the precession period will slightly decrease, while the rotation period ranges between $\simeq 230$ and $\simeq 290$~hours, roughly centered at the present value. There is also a clear correlation between $P_\phi$ and $P_\psi$ demonstrated on the left panel of Fig.~\ref{fig:periods_nom}. These findings are similar but quite more accurate than those presented in \citet{2023Icar..39015324B}.

Figure~\ref{fig:pint_nom} shows pre- and post-encounter values of the dimensionless parameter $p=2EB/L^2$, diagnostic about the tumbling mode.
All values are smaller than unity, implying the post-encounter tumbling will remain in the SAM mode in this set of simulations. The right panel of this figure indicates that the mean effect of Earth's gravitational torque during the encounter shifts the rotation state towards the separatrix between the SAM and LAM modes.

Finally, Fig.~\ref{fig:pole_nom} shows the predicted change in orientation of the rotational angular momentum vector $\vec{L}$ in the inertial space. Prior to the effect displayed in this figure, which is entirely due to the Earth encounter, we note a steady (regular) precession due to the solar gravitational torque that amounts to about $3^\circ$ in $8$~years.%
\footnote{This value is well expected and roughly corresponds to the pole precession of (2100) Ra-Shalom on a similar heliocentric orbit and having a similarly slow rotation \citep[see][]{2024A&A...682A..93D}.}
This effect has not been taken into account when fitting the available observations in Sec.~\ref{sec:model}, but this is justified by more than $10^\circ$ uncertainty in the longitude of $\vec{L}$. More than a $100^\circ$ change in pole longitude during the Earth encounter appears dramatically large, but most of the effect is apparent. Due to near-polar latitude, the real tilt of $\vec{L}$ is much smaller. Its median value amounts to $\simeq 9.8^\circ$ only.

Taking into account the slight change in the orbital inclination, the median change in Apophis obliquity $\epsilon$ (defined here as the angle between the normal to the osculating orbit and the rotational angular momentum $\vec{L}$) is only $\simeq 5.7^\circ$. The fractional contribution to the change in the Yarkovsky acceleration is then estimated to $\simeq 0.9$\%. 
This is formally larger than the current accuracy of the Yarkovsky effect determination for Apophis \citep[about $0.5$\%; e.g.,][]{ 2022LPICo2681.2007F,2024LPICo3006.2046F}. However, the post-encounter value of the Yarkovsky drift rate in semimajor axis $\dd a / \dd t$ is affected by more parameters than just the obliquity value. It turns out that the dominant contribution is due to the change in the semimajor axis, followed by a smaller contribution due to the rotation period. While detailed analysis should await until 2029, we note that the scaling $\dd a/\dd t\propto a^{-1/2}\cos\epsilon$ made by \citet{2023Icar..39015324B} is not correct. This is because it does not take into account the role of the thermal parameter $\Theta$ \citep[see, e.g.,][]{Bot.ea:06,Vok.ea:15}. The slow rotation rate and high subsolar temperature imply $\Theta< 1$, unless very high value of the surface thermal inertia $\Gamma$. For instance, assuming $\Gamma\simeq 500$ [SI units] \citep[roughly at the upper limit of data reported in][]{2014A&A...566A..22M,2022PSJ.....3..124S,2023PSJ.....4..198D}, we have $\Theta\simeq 0.5$. In this regime $\dd a/\dd t\propto \Theta$, so an increase of $\Theta$ due to an increase of the semimajor axis and a slight decrease of the rotation period $P_\phi$ will increase $\dd a/\dd t$ value. Considering the corresponding scaling $\dd a/\dd t\propto a P_\phi^{-1/2}\cos\epsilon$, we may expect about $20$\% increase of the Yarkovsky effect from the analysis in this Section (things might be modified if more significant changes in $P_\phi$ and $\epsilon$ apply as in some simulations with the YORP torque discussed in the next Section).

\subsubsection{Potential role of radiative torques}\label{spin_YORP}
Next, we extend our simulations from the previous Section by the effects of the YORP torque (\ref{e8}). Unlike the gravitational torques (\ref{e7}), which depend only on the dimensionless fractions $\alpha=B/A$ and $\beta=B/C$ defined by the principal values of the inertia tensor, whose solution is part of the lightcurve inversion in Sec.~\ref{sec:model}, for the YORP torque, we have to define more parameters. We use the triangulated shapes from the lightcurve inversion for each of the 1000 bootstrapped variants of the solution A (having typically 900 to 1000 surface vertices). We assume uniform density of $2$~g~cm$^{-3}$ and scale each of the models to a volume equivalent to a sphere of $400$\,m diameter, which is a rounded value of the diameter 417\,m found by \cite{Bro.ea:26} when scaling our Apophis shape model by occultations.
For the sake of an illustration, we use a fudge factor $f=0.4$ in (\ref{e8}). This value is a typical reduction by which the detected YORP values differ from the prediction of simple models like (\ref{e8}). We also ran simulations with smaller and larger values of $f$, but without observational constraints on the YORP effect for Apophis, we report their results only briefly. More efforts in this respect are postponed to further Apophis spin state models derived from more observational data (Sec.~\ref{concl}).

Figures~\ref{fig:periods_yorp} to \ref{fig:pole_yorp} show again the change of the principal spin parameters by comparing their pre- and post-encounter values or distributions: (i) $P_\phi$ and $P_\psi$ periods (Fig.~\ref{fig:periods_yorp}), (ii) $p$ parameter (Fig.~\ref{fig:pint_yorp}), and (iii) direction of the rotational angular momentum vector $\vec{L}$ (Fig.~\ref{fig:pole_yorp}). The $p$ and $\vec{L}$ change is markedly larger than in Sec.~\ref{spin_nominal}, where only gravitational torques were considered, especially $p$ values now have a tail towards unity (SAM to LAM separatrix limit). Nevertheless, in all tested variants of model A, the post-encounter tumbling state remained in the SAM mode for this value of the fudge factor \citep[simulations with the fudge factor $f\geq 0.7$ showed a small fraction of LAM post-encounter states, see also][]{2023Icar..39015324B}. However, the main difference concerns the precession and rotation values. Their post-encounter values now exhibit a much larger spread.

\section{Conclusions} \label{concl}
The photometric dataset of Apophis obtained during its two favorable apparitions in 2012--2013 and 2020--2021 is abundant, yet insufficient to uniquely determine the two periods of Apophis's excited rotation. Due to the long interval between the apparitions not covered by the data, there are at least three different combinations of the precession and rotation periods that all produce about the same synthetic light curves and thus the same fit to the data. The possibility to constrain Apophis's post-encounter state after April 2029, important among other topics for its future orbital evolution, including analysis of the Earth impact hazard \citep[e.g.,][]{Vok.apophis:15}, is the primary motivation for a unique and accurate pre-encounter model of its spin-state. This is because the resulting effect of Earth's gravitational torque depends very sensitively on the exact phase of Apophis's rotation state at the moment of the closest encounter.

In this work, we found that despite numerous and accurate past observations, neither uniqueness nor precision has yet been achieved. There are at least three statistically equivalent groups of solutions for Apophis's spin state at this moment, and their prediction of its pre-encounter orientation is very similar, within $20^\circ$ in terms of Euler angles. This is because the interval between the last apparition covered by the data and 2029 is eight years, which is the same interval as that between the observing windows in 2012--2013 and 2020--2021. Therefore, we can predict the approximate orientation of Apophis during its close encounter, but its accuracy is insufficient to reliably predict its post-encounter spin state so far. The simulations that include only the effect of Earth's gravitational torques show that (i) the precession period will change from the current value of $P_\phi = 27.374 \pm 0.001$\,h towards shorter values between 24.5--27.5\,h, and (ii) the rotation period will change from $P_\psi = 262.2 \pm 0.1$\,h to the interval 230--290\,h. At the same time, the direction of the rotational angular momentum vector will tilt by about $10^\circ$, and the rotation will stay in the SAM mode. When the so far unconstrained YORP torque is included in the model, the uncertainty of the prediction is much larger, with the post-encounter $P_\phi$ ranging from 16 to 35\,h and $P_\psi$ from 160 to 400\,h. 

Because the apparition in 2029 is not suitable for breaking the degeneracy in the rotation and precession periods, it is crucial to collect some photometric data before 2029. There will be an opportunity to take critical data to distinguish between the several possible models of Apophis's spin state in 2027 and 2028.  The observational conditions and requirements were discussed by \cite{Pra.Dur:24}.  Summarizing their findings briefly, we note that there will be observational windows for Apophis photometry from February to April 2027 and from December 2027 to June 2028 when Apophis will be $>60^\circ$ from Sun.  The observing conditions will be particularly challenging in the February--April 2027 window when Apophis's mean $V$ is predicted to be 21.1--21.5\,mag and large (4+m) telescopes in the northern hemisphere will be needed.  The observing conditions will be slightly more favorable in the December 2027--June 2028 window, with Apophis's mean $V$ ranging from 19.2 to 20.7\,mag; medium-sized (2--4m) telescopes in both hemispheres will be required.  However, data taken in both observational windows may be needed not only to distinguish between the several possible Apophis's spin models, but also to estimate or constrain the YORP effect in the asteroid.  We encourage photometric observers to take the needed high-quality and extensive (though with a relatively low cadence) photometric data for Apophis in both apparitions.
  
\begin{acknowledgements}
  This work was supported by the grant 23-04946S of the Czech Science Foundation. P.P.~has been supported by the {\it Praemium Academiae} award (no. AP2401) from the
Academy of Sciences of the Czech Republic.
  This publication uses data products from the TRAPPIST project funded by the Belgian National Fund for Scientific Research (F.R.S.-FNRS) under grant PDR T.0120.21. TRAPPISTNorth is funded by the University of Liege in collaboration with Cadi Ayyad University of Marrakech. Data were collected at the ESO La Silla Observatory. EJ is the director of research at FNRS. 
  Based on observations obtained with PESTO at the Mont-Mégantic Observatory, funded by the Université de Montréal, Université Laval, the Natural Sciences and Engineering Research Council of Canada (NSERC), the Fonds québécois de la recherche sur la Nature et les technologies (FQRNT), and the Canada Economic Development program. This paper was partially based on observations obtained at the Optical Wide-field Patrol Network (OWL-Net), the Sobaeksan Optical Astronomy Observatory (SOAO), the Lemmonsan Optical Astronomy Observatory (LOAO), the Bohyunsan Optical Astronomy Observatory (BOAO), and the Korea Microlensing Telescope Network (KMTNet), all of which are operated by the Korea Astronomy and Space Science Institute (KASI).
\end{acknowledgements}

\bibliographystyle{aa}
\bibliography{bibliography_joined}

@preamble{ "\newcommand{\SortNoop}[1]{} " }

@INPROCEEDINGS{Ara.ea:25,
       author = {{Arai}, T. and {Destiny+ Team}},
        title = "{Planetary Defense and the DESTINY+ Mission}",
    booktitle = {Apophis T-4 Years},
         year = 2025,
       series = {LPI Contributions},
       volume = {3083},
        month = apr,
          eid = {2061},
        pages = {2061},
       adsurl = {https://ui.adsabs.harvard.edu/abs/2025LPICo3083.2061A}
}

@ARTICLE{Bin.ea:09,
       author = {{Binzel}, Richard P. and {Rivkin}, Andrew S. and {Thomas}, Cristina A. and {Vernazza}, Pierre and {Burbine}, Thomas H. and {DeMeo}, Francesca E. and {Bus}, Schelte J. and {Tokunaga}, Alan T. and {Birlan}, Mirel},
        title = "{Spectral properties and composition of potentially hazardous Asteroid (99942) Apophis}",
      journal = {\icarus},
         year = 2009,
        month = apr,
       volume = {200},
       number = {2},
        pages = {480-485},
          doi = {10.1016/j.icarus.2008.11.028},
       adsurl = {https://ui.adsabs.harvard.edu/abs/2009Icar..200..480B}
}

@ARTICLE{Bot.ea:06,
   author = {{Bottke}, Jr., W.~F. and {Vokrouhlick{\'y}}, D. and {Rubincam}, D.~P. and
	{Nesvorn{\'y}}, D.},
    title = "{The Yarkovsky and Yorp effects: Implications for asteroid dynamics}",
  journal = {Annual Review of Earth and Planetary Sciences},
    year = 2006,
    month = may,
   volume = 34,
    pages = {157-191},
      doi = {10.1146/annurev.earth.34.031405.125154PDF: http://arjournals.annualreviews.org/doi/pdf/10.1146/annurev.earth.34.031405.125154},
   adsurl = {http://adsabs.harvard.edu/cgi-bin/nph-bib_query?bibcode=2006AREPS..34..157B&db_key=AST}
}

@ARTICLE{Bre.ea:09,
   author = {{Breiter}, S. and {Bartczak}, P. and {Czekaj}, M. and {Oczujda}, B. and 
	{Vokrouhlick{\'y}}, D.},
    title = "{The YORP effect on 25 143 Itokawa}",
  journal = {\aap},
     year = 2009,
    month = nov,
   volume = 507,
    pages = {1073-1081},
      doi = {10.1051/0004-6361/200912543},
   adsurl = {http://adsabs.harvard.edu/abs/2009A%26A...507.1073B}
}

@ARTICLE{Bre.ea:11,
       author = {{Breiter}, S. and {Ro{\.z}ek}, A. and {Vokrouhlick{\'y}}, D.},
        title = "{Yarkovsky-O'Keefe-Radzievskii-Paddack effect on tumbling objects}",
      journal = {\mnras},
         year = 2011,
        month = nov,
       volume = {417},
       number = {4},
        pages = {2478-2499},
          doi = {10.1111/j.1365-2966.2011.19411.x},
       adsurl = {https://ui.adsabs.harvard.edu/abs/2011MNRAS.417.2478B}
}

@ARTICLE{Bro.ea:26,
       author = {{Bro{\v{z}}}, M. and {Binzel}, R.~P. and {Vernazza}, P. and {Marsset}, M. and {Chrenko}, O. and {{\v{D}}urech}, J. and {Herald}, D.},
        title = "{Apophis source population and Earth encounter frequency of Apophis-like bodies}",
      journal = {\aap},
         year = 2026,
        month = apr,
       volume = {708},
          eid = {A162},
        pages = {A162},
          doi = {10.1051/0004-6361/202557980},
archivePrefix = {arXiv},
       eprint = {2602.19849},
 primaryClass = {astro-ph.EP},
       adsurl = {https://ui.adsabs.harvard.edu/abs/2026A&A...708A.162B}
}

@ARTICLE{Bro.ea:18,
       author = {{Brozovi{\'c}}, Marina and {Benner}, Lance A.~M. and {McMichael}, Joseph G. and {Giorgini}, Jon D. and {Pravec}, Petr and {Scheirich}, Petr and {Magri}, Christopher and {Busch}, Michael W. and {Jao}, Joseph S. and {Lee}, Clement G. and {Snedeker}, Lawrence G. and {Silva}, Marc A. and {Slade}, Martin A. and {Semenov}, Boris and {Nolan}, Michael C. and {Taylor}, Patrick A. and {Howell}, Ellen S. and {Lawrence}, Kenneth J.},
        title = "{Goldstone and Arecibo radar observations of (99942) Apophis in 2012-2013}",
      journal = {\icarus},
         year = 2018,
        month = jan,
       volume = {300},
        pages = {115-128},
          doi = {10.1016/j.icarus.2017.08.032},
       adsurl = {https://ui.adsabs.harvard.edu/abs/2018Icar..300..115B}
}

@ARTICLE{Cap.Vok:04,
    author = {{\SortNoop{Capek1}{\v C}apek}, D. and {Vokrouhlick{\' y}}, D.},
    title = "{The YORP effect with finite thermal conductivity}",
    journal = {Icarus},
    year = 2004,
    month = dec,
    volume = 172,
    pages = {526-536},
    doi = {10.1016/j.icarus.2004.07.003},
    adsurl = {http://adsabs.harvard.edu/abs/2004Icar..172..526C},
}

@ARTICLE{Cic.Sch:10,
       author = {{Cical{\`o}}, S. and {Scheeres}, D.~J.},
        title = "{Averaged rotational dynamics of an asteroid in tumbling rotation under the YORP torque}",
      journal = {Celestial Mechanics and Dynamical Astronomy},
         year = 2010,
        month = apr,
       volume = {106},
       number = {4},
        pages = {301-337},
          doi = {10.1007/s10569-009-9249-7},
       adsurl = {https://ui.adsabs.harvard.edu/abs/2010CeMDA.106..301C}
}

@ARTICLE{Dep.Eli:93,
       author = {{Deprit}, Andre and {Elipe}, Antonio},
        title = "{Complete reduction of the Euler-Poinsot problem}",
      journal = {Journal of the Astronautical Sciences},
         year = 1993,
        month = oct,
       volume = {41},
       number = {4},
        pages = {603-628},
       adsurl = {https://ui.adsabs.harvard.edu/abs/1993JAnSc..41..603D}
}

@ARTICLE{Dob:96,
    author = {{Dobrovolskis}, A.~R.},
    title = "{Inertia of any polyhedron}",
    journal = {Icarus},
    year = 1996,
    month = dec,
    volume = 124,
    pages = {698-704},
    adsurl = {http://adsabs.harvard.edu/cgi-bin/nph-bib_query?bibcode=1996Icar..124..698D&db_key=AST}
}

@article{Far.ea:13b,
title = {Yarkovsky-driven impact risk analysis for asteroid (99942) Apophis},
journal = {Icarus},
volume = {224},
number = {1},
pages = {192-200},
year = {2013},
issn = {0019-1035},
doi = {https://doi.org/10.1016/j.icarus.2013.02.020},
url = {https://www.sciencedirect.com/science/article/pii/S0019103513000821},
author = {D. Farnocchia and S.R. Chesley and P.W. Chodas and M. Micheli and D.J. Tholen and A. Milani and G.T. Elliott and F. Bernardi},
}

@ARTICLE{Fat.ea:25,
       author = {{Fatka}, P. and {Pravec}, P. and {Scheirich}, P. and {Ku{\v{s}}nir{\'a}k}, P. and {Hornoch}, K. and {Ku{\v{c}}{\'a}kov{\'a}}, H. and {Ergashev}, K.~E. and {Souza de Joode}, M. and {Burkhonov}, O.~A. and {Ehgamberdiev}, S.~A. and {Gal{\'a}d}, A. and {Vil{\'a}gi}, J. and {Reddy}, V. and {Dyvig}, R. and {Ries}, J.~G. and {Snodgrass}, C. and {Donaldson}, A. and {Peixinho}, N. and {Khalouei}, E.},
        title = "{Spins and shapes of 11 near-Earth asteroids observed within the NEOROCKS project}",
      journal = {\aap},
         year = 2025,
        month = mar,
       volume = {695},
          eid = {A139},
        pages = {A139},
          doi = {10.1051/0004-6361/202450027},
       adsurl = {https://ui.adsabs.harvard.edu/abs/2025A&A...695A.139F}
}

@BOOK{Fit:70,
  title={Principles of Celestial Mechanics},
  author={Fitzpatrick, P.M.},
  year={1970},
  publisher={Academic Press, New York}
}

@ARTICLE{Gol.Kru:12,
   author = {{Golubov}, O. and {Krugly}, Y.~N.},
    title = "{Tangential component of the YORP effect}",
  journal = {\apjl},
     year = 2012,
    month = jun,
   volume = 752,
      eid = {L11},
    pages = {L11},
      doi = {10.1088/2041-8205/752/1/L11},
   adsurl = {http://adsabs.harvard.edu/abs/2012ApJ...752L..11G}
}

@ARTICLE{Hap:86,
    author = {{Hapke}, B.},
    title = "{Bidirectional reflectance spectroscopy. IV -- The extinction coefficient and the opposition effect}",
    journal = {Icarus},
    year = 1986,
    month = aug,
    volume = 67,
    pages = {264-280},
    adsurl = {http://adsabs.harvard.edu/cgi-bin/nph-bib_query?bibcode=1986Icar...67..264H&db_key=AST}
}

@BOOK{Hap:12,
   author = {{Hapke}, B.},
    title = "{Theory of Reflectance and Emittance Spectroscopy}",
     year = 2012,
publisher = {Cambridge University Press}
}

@INCOLLECTION{Hel.Vev:89,
   author = {{Helfenstein}, P. and {Veverka}, J.},
   title = "{Physical characterization of asteroid surfaces from photometric analysis}", 
   booktitle = "{Asteroids~II}",
   publisher = "{University of Arizona Press}",
   address = {Tucson},
   year = 1989,
   editor = {{Binzel}, R.~P. and {Gehrels}, T. and {Matthews}, M.~S.},
   pages = {557-593}
}

@ARTICLE{Kaa.Tor:01,
    author = {{Kaasalainen}, M. and {Torppa}, J.},
    title = "{Optimization methods for asteroid lightcurve inversion. I. Shape determination}",
    journal = {Icarus},
    year = 2001,
    month = sep,
    volume = 153,
    pages = {24-36},
    adsurl = {http://adsabs.harvard.edu/cgi-bin/nph-bib_query?bibcode=2001Icar..153...24K&db_key=AST}
}

@ARTICLE{Kaa:01,
    author = {{Kaasalainen}, M.},
    title = "{Interpretation of lightcurves of precessing asteroids}",
    journal = {Astron. Astrophys.},
    year = 2001,
    month = sep,
    volume = 376,
    pages = {302-309},
    adsurl = {http://adsabs.harvard.edu/cgi-bin/nph-bib_query?bibcode=2001A%26A...376..302K&db_key=AST}
}

@BOOK{Lan.Lif:69,
    author = {Landau, L.D. and Lifshitz, E.M.},
    title = "{Mechanics}",
    publisher = {Pergamon Press, Oxford}, 
    year = {1969}
}

@ARTICLE{Lan:92,
       author = {{Landolt}, Arlo U.},
        title = "{UBVRI Photometric Standard Stars in the Magnitude Range 11.5 < V < 16.0 Around the Celestial Equator}",
      journal = {\aj},
         year = 1992,
        month = jul,
       volume = {104},
        pages = {340},
          doi = {10.1086/116242},
       adsurl = {https://ui.adsabs.harvard.edu/abs/1992AJ....104..340L}
}

@INPROCEEDINGS{Laz.ea:25,
       author = {{Lazzarin}, Monica and {Michel}, Patrick and {Kueppers}, Michael and {Green}, Simon and {Tortora}, Paolo and {Ulamec}, Stephan and {Baptiste Vincent}, Jean and {Abell}, Paul and {Sugita}, Seiji and {Martino}, Paolo},
        title = "{RAMSES: A European rendezvous mission to study tidal effects on the Near-Earth Asteroid Apophis during its 2029 close encounter with the Earth}",
    booktitle = {EPSC-DPS Joint Meeting 2025},
         year = 2025,
       volume = {2025},
        month = sep,
          eid = {EPSC-DPS2025-806},
        pages = {EPSC-DPS2025-806},
          doi = {10.5194/epsc-dps2025-806},
       adsurl = {https://ui.adsabs.harvard.edu/abs/2025epsc.conf..806L}
}

@ARTICLE{Lee.ea:21,
       author = {{Lee}, Hee-Jae and {{\v{D}}urech}, Josef and {Vokrouhlick{\'y}}, David and {Pravec}, Petr and {Moon}, Hong-Kyu and {Ryan}, William and {Kim}, Myung-Jin and {Kim}, Chun-Hwey and {Choi}, Young-Jun and {Bacci}, Paolo and {Pollock}, Joe and {Apitzsch}, Rolf},
        title = "{Spin Change of Asteroid 2012 TC4 Probably by Radiation Torques}",
      journal = {\aj},
         year = 2021,
        month = mar,
       volume = {161},
       number = {3},
          eid = {112},
        pages = {112},
          doi = {10.3847/1538-3881/abd4da},
archivePrefix = {arXiv},
       eprint = {2012.08771},
 primaryClass = {astro-ph.EP},
       adsurl = {https://ui.adsabs.harvard.edu/abs/2021AJ....161..112L}
}

@ARTICLE{Lee.ea:22,
       author = {{Lee}, H. -J. and {Kim}, M. -J. and {Marciniak}, A. and {Kim}, D. -H. and {Moon}, H. -K. and {Choi}, Y. -J. and {Zo{\l}a}, S. and {Chatelain}, J. and {Lister}, T.~A. and {Gomez}, E. and {Greenstreet}, S. and {P{\'a}l}, A. and {Szak{\'a}ts}, R. and {Erasmus}, N. and {Lees}, R. and {Janse van Rensburg}, P. and {Og{\l}oza}, W. and {Dr{\'o}{\.z}d{\.z}}, M. and {{\.Z}ejmo}, M. and {Kami{\'n}ski}, K. and {Kami{\'n}ska}, M.~K. and {Duffard}, R. and {Roh}, D. -G. and {Yim}, H. -S. and {Kim}, T. and {Mottola}, S. and {Yoshida}, F. and {Reichart}, D.~E. and {Sonbas}, E. and {Caton}, D.~B. and {Kaplan}, M. and {Erece}, O. and {Yang}, H.},
        title = "{Refinement of the convex shape model and tumbling spin state of (99942) Apophis using the 2020-2021 apparition data}",
      journal = {\aap},
         year = 2022,
        month = may,
       volume = {661},
          eid = {L3},
        pages = {L3},
          doi = {10.1051/0004-6361/202243442},
archivePrefix = {arXiv},
       eprint = {2204.02540},
 primaryClass = {astro-ph.EP},
       adsurl = {https://ui.adsabs.harvard.edu/abs/2022A&A...661L...3L}
}

@INCOLLECTION{Li.ea:15,
   author = {{Li}, J.-Y. and {Helfenstein}, P. and {Burrati}, B.~J. and {Takir}, D. and {Clark}, B.~E.},
   title = "{Asteroid photometry}", 
   booktitle = "{Asteroids~IV}",
   publisher = "{University of Arizona Press}",
   address = {Tucson},
   pages = {129-150},
   year = 2015,
   editor = {{Michel}, P. and {DeMeo}, F.~E. and {Bottke}, W.~F.}
}

@article{Lic.ea:16,
    title={GTC/CanariCam observations of (99942) Apophis},
    author={J. Licandro and T. Müller and C. Álvarez and V. Alí-Lagoa and M. Delbo’},
    journal={Astronomy and Astrophysics},
    year={2016},
    volume={585},
    pages={1-4},
    doi={10.1051/0004-6361/201526888}
}

@INPROCEEDINGS{Mic.ea:25,
       author = {{Michel}, P. and {Lazzazin}, M. and {K{\"u}ppers}, M. and {Green}, S. and {Tortora}, P. and {Ulamec}, S. and {Vincent}, J.~B. and {Abell}, P. and {Sugita}, S. and {Martino}, P.},
        title = "{Science Objectives of RAMSES: ESA's Rapid Apophis Mission for SpacE Safety}",
    booktitle = {Apophis T-4 Years},
         year = 2025,
       series = {LPI Contributions},
       volume = {3083},
        month = apr,
          eid = {2012},
        pages = {2012},
       adsurl = {https://ui.adsabs.harvard.edu/abs/2025LPICo3083.2012M}
}

@INPROCEEDINGS{Nol.ea:25,
       author = {{Nolan}, Michael C. and {DellaGiustina}, Daniella N. and {Polit}, Anjani T. and {Golish}, Dathon R. and {Moreau}, Michael C. and {Simon}, Amy A. and {Guzewich}, Scott D.},
        title = "{OSIRIS-APEX: NASA's Apophis Explorer Mission}",
    booktitle = {EPSC-DPS Joint Meeting 2025},
         year = 2025,
       volume = {2025},
        month = sep,
          eid = {EPSC-DPS2025-187},
        pages = {EPSC-DPS2025-187},
          doi = {10.5194/epsc-dps2025-187},
       adsurl = {https://ui.adsabs.harvard.edu/abs/2025epsc.conf..187N}
}

@ARTICLE{Pra.ea:14,
   author = {{Pravec}, P. and {Scheirich}, P. and {{\v D}urech}, J. and {Pollock}, J. and 
	{Ku{\v s}nir{\'a}k}, P. and {Hornoch}, K. and {Gal{\'a}d}, A. and 
	{Vokrouhlick{\'y}}, D. and {Harris}, A.~W. and {Jehin}, E. and 
	{Manfroid}, J. and {Opitom}, C. and {Gillon}, M. and {Colas}, F. and 
	{Oey}, J. and {Vra{\v s}til}, J. and {Reichart}, D. and {Ivarsen}, K. and 
	{Haislip}, J. and {LaCluyze}, A.},
    title = "{The tumbling spin state of (99942) Apophis}",
  journal = {\icarus},
     year = 2014,
    month = may,
   volume = 233,
    pages = {48-60},
      doi = {10.1016/j.icarus.2014.01.026},
   adsurl = {http://adsabs.harvard.edu/abs/2014Icar..233...48P}
}

@INPROCEEDINGS{Pra.Dur:24,
       author = {{Pravec}, P. and {\v{D}urech}, J.},
        title = "{Apophis Observational Opportunities from Now until the 2029 April 13 Close Approach, to Improve Solution for its Spin State}",
    booktitle = {Apophis T-5 Workshop},
         year = 2024,
       series = {LPI Contributions},
       volume = {3006},
        month = apr,
          eid = {2008},
        pages = {2008},
       adsurl = {https://ui.adsabs.harvard.edu/abs/2024LPICo3006.2008P}
}

@ARTICLE{Rub:00,
   author = {{Rubincam}, D.~P.},
    title = "{Radiative spin-up and spin-down of small asteroids}",
  journal = {Icarus},
     year = 2000,
    month = nov,
   volume = 148,
    pages = {2-11},
      doi = {10.1006/icar.2000.6485},
   adsurl = {http://adsabs.harvard.edu/abs/2000Icar..148....2R}
}

@ARTICLE{Sch.ea:07,
   author = {{Scheeres}, D.~J. and {Abe}, M. and {Yoshikawa}, M. and {Nakamura}, R. and 
	{Gaskell}, R.~W. and {Abell}, P.~A.},
    title = "{The effect of YORP on Itokawa}",
  journal = {Icarus},
     year = 2007,
    month = jun,
   volume = 188,
    pages = {425-429},
      doi = {10.1016/j.icarus.2006.12.014},
   adsurl = {http://adsabs.harvard.edu/abs/2007Icar..188..425S}
}

@ARTICLE{Tak.ea:13,
       author = {{Takahashi}, Yu and {Busch}, Michael W. and {Scheeres}, D.~J.},
        title = "{Spin State and Moment of Inertia Characterization of 4179 Toutatis}",
      journal = {\aj},
         year = 2013,
        month = oct,
       volume = {146},
       number = {4},
          eid = {95},
        pages = {95},
          doi = {10.1088/0004-6256/146/4/95},
       adsurl = {https://ui.adsabs.harvard.edu/abs/2013AJ....146...95T}
}

@ARTICLE{Vok.Cap:02,
    author = {{Vokrouhlick{\'y}}, D. and {{\v C}apek}, D.},
    title = "{YORP-Induced long-term evolution of the spin state of small asteroids and meteoroids: Rubincam's approximation}",
    journal = {Icarus},
    year = 2002,
    month = oct,
    volume = 159,
    pages = {449-467},
    adsurl = {http://adsabs.harvard.edu/cgi-bin/nph-bib_query?bibcode=2002Icar..159..449V&amp;db_key=AST}
}

@ARTICLE{Vok.ea:07,
   author = {{Vokrouhlick{\'y}}, D. and {Breiter}, S. and {Nesvorn{\'y}}, D. and 
	{Bottke}, W.~F.},
    title = "{Generalized YORP evolution: Onset of tumbling and new asymptotic states}",
  journal = {Icarus},
     year = 2007,
    month = nov,
   volume = 191,
    pages = {636-650},
      doi = {10.1016/j.icarus.2007.06.002},
   adsurl = {http://adsabs.harvard.edu/abs/2007Icar..191..636V}
}

@ARTICLE{Vok.apophis:15,
       author = {{Vokrouhlick\'y}, David and {Farnocchia}, Davide and
         {{\v{C}}apek}, David and {Chesley}, Steven R. and {Pravec}, Petr and
         {Scheirich}, Petr and {M{\"u}ller}, Thomas G.},
        title = "{The Yarkovsky effect for 99942 Apophis}",
      journal = {\icarus},
         year = 2015,
        month = may,
       volume = {252},
        pages = {277-283},
          doi = {10.1016/j.icarus.2015.01.011},
       adsurl = {https://ui.adsabs.harvard.edu/abs/2015Icar..252..277V}
}

@INCOLLECTION{Vok.ea:15,
   author = {{Vokrouhlick\'y}, D. and {Bottke}, W.~F. and {Chesley}, S.~R. and {Scheeres}, D.~J. and {Statler}, T.~S.},
   title = "{The Yarkovsky and YORP effects}", 
   booktitle = "{Asteroids~IV}",
   publisher = "{University of Arizona Press}",
   address = {Tucson},
   pages = {509-531},
   year = 2015,
   editor = {{Michel}, P. and {DeMeo}, F.~E. and {Bottke}, W.~F.}
}

@BOOK{Whi:1917,
  title={A Treatise on the Analytical Dynamics of Particles and Rigid Bodies},
  author={Whittaker, E.T.},
  year={1917},
  publisher={Cambridge University Press, Cambridge}
}

@ARTICLE{Yu.ea:14,
       author = {{Yu}, Yang and {Richardson}, Derek C. and {Michel}, Patrick and {Schwartz}, Stephen R. and {Ballouz}, Ronald-Louis},
        title = "{Numerical predictions of surface effects during the 2029 close approach of Asteroid 99942 Apophis}",
      journal = {\icarus},
         year = 2014,
        month = nov,
       volume = {242},
        pages = {82-96},
          doi = {10.1016/j.icarus.2014.07.027},
archivePrefix = {arXiv},
       eprint = {1408.0168},
 primaryClass = {astro-ph.EP},
       adsurl = {https://ui.adsabs.harvard.edu/abs/2014Icar..242...82Y}
}

@ARTICLE{2024PSJ.....5..251B,
       author = {{Ballouz}, R. -L. and {Agrusa}, H. and {Barnouin}, O.~S. and {Walsh}, K.~J. and {Zhang}, Y. and {Binzel}, R.~P. and {Bray}, V.~J. and {DellaGiustina}, D.~N. and {Jawin}, E.~R. and {DeMartini}, J.~V. and {Marusiak}, A. and {Michel}, P. and {Murdoch}, N. and {Richardson}, D.~C. and {Rivera-Valent{\'\i}n}, E.~G. and {Rivkin}, A.~S. and {Tang}, Y.},
        title = "{Shaking and Tumbling: Short- and Long-timescale Mechanisms for Resurfacing of Near-Earth Asteroid Surfaces from Planetary Tides and Predictions for the 2029 Earth Encounter by (99942) Apophis}",
      journal = {\psj},
         year = 2024,
        month = nov,
       volume = {5},
       number = {11},
          eid = {251},
        pages = {251},
          doi = {10.3847/PSJ/ad84f2},
archivePrefix = {arXiv},
       eprint = {2406.04864},
 primaryClass = {astro-ph.EP},
       adsurl = {https://ui.adsabs.harvard.edu/abs/2024PSJ.....5..251B}
}

@INPROCEEDINGS{2024LPICo3006.2060B,
       author = {{Ballouz}, R. -L. and {Graninger}, D.~M. and {Adams}, E.~Y. and {Atchison}, J.~A. and {Barnouin}, O.~S. and {Berdis}, J. and {Bull}, R. and {Chabot}, N.~L. and {Daly}, R.~T. and {Davis}, A.~K. and {Ernst}, C.~M. and {Rivkin}, A.~S. and {Shannon}, J. and {Siddique}, F.~E. and {Srinivasan}, D.},
        title = "{Flyby Asteroid Reconnaissance (FLARE) Mission to Apophis: A Mission Concept to Apophis before its Earth Encounter to Demonstrate Flyby Reconnaissance for Planetary Defense}",
    booktitle = {Apophis T-5 Workshop},
         year = 2024,
       series = {LPI Contributions},
       volume = {3006},
        month = apr,
          eid = {2060},
        pages = {2060},
       adsurl = {https://ui.adsabs.harvard.edu/abs/2024LPICo3006.2060B}
}

@ARTICLE{2023Icar..39015324B,
       author = {{Benson}, Conor J. and {Scheeres}, Daniel J. and {Brozovi{\'c}}, Marina and {Chesley}, Steven R. and {Pravec}, Petr and {Scheirich}, Petr},
        title = "{Spin state evolution of (99942) Apophis during its 2029 Earth encounter}",
      journal = {\icarus},
         year = 2023,
        month = jan,
       volume = {390},
          eid = {115324},
        pages = {115324},
          doi = {10.1016/j.icarus.2022.115324},
archivePrefix = {arXiv},
       eprint = {2210.13365},
 primaryClass = {astro-ph.EP},
       adsurl = {https://ui.adsabs.harvard.edu/abs/2023Icar..39015324B}
}

@INPROCEEDINGS{2022LPICo2681.2023B,
       author = {{Brozovic}, M. and {Benner}, L.~A.~M. and {Naidu}, S.~P. and {Busch}, M.~W. and {Giorgini}, J.~D. and {Lazio}, J. and {Hall}, T.},
        title = "{Radar Observations of 99942 Apophis in 2021 and Plans for 2029}",
    booktitle = {Apophis T-7 Years: Knowledge Opportunities for the Science of Planetary Defense},
         year = 2022,
       series = {LPI Contributions},
       volume = {2681},
        month = may,
          eid = {2023},
        pages = {2023},
       adsurl = {https://ui.adsabs.harvard.edu/abs/2022LPICo2681.2023B}
}

@ARTICLE{2023PSJ.....4..198D,
       author = {{DellaGiustina}, Daniella N. and {Nolan}, Michael C. and {Polit}, Anjani T. and {Moreau}, Michael C. and {Golish}, Dathon R. and {Simon}, Amy A. and {Adam}, Coralie D. and {Antreasian}, Peter G. and {Ballouz}, Ronald-Louis and {Barnouin}, Olivier S. and {Becker}, Kris J. and {Bennett}, Carina A. and {Binzel}, Richard P. and {Bos}, Brent J. and {Burns}, Richard and {Castro}, Nayessda and {Chesley}, Steven R. and {Christensen}, Philip R. and {Crombie}, M. Katherine and {Daly}, Michael G. and {Daly}, R. Terik and {Enos}, Heather L. and {Farnocchia}, Davide and {Freund Kasper}, Sandra and {Garcia}, Rose and {Getzandanner}, Kenneth M. and {Guzewich}, Scott D. and {Haberle}, Christopher W. and {Haltigin}, Timothy and {Hamilton}, Victoria E. and {Harshman}, Karl and {Hatten}, Noble and {Hughes}, Kyle M. and {Jawin}, Erica R. and {Kaplan}, Hannah H. and {Lauretta}, Dante S. and {Leonard}, Jason M. and {Levine}, Andrew H. and {Liounis}, Andrew J. and {May}, Christian W. and {Mayorga}, Laura C. and {Nguyen}, Lillian and {Quick}, Lynnae C. and {Reuter}, Dennis C. and {Rivera-Valent{\'\i}n}, Edgard and {Rizk}, Bashar and {Roper}, Heather L. and {Ryan}, Andrew J. and {Sutter}, Brian and {Westermann}, Mathilde M. and {Wibben}, Daniel R. and {Williams}, Bobby G. and {Williams}, Kenneth and {Wolner}, C.~W.~V.},
        title = "{OSIRIS-APEX: An OSIRIS-REx Extended Mission to Asteroid Apophis}",
      journal = {\psj},
         year = 2023,
        month = oct,
       volume = {4},
       number = {10},
          eid = {198},
        pages = {198},
          doi = {10.3847/PSJ/acf75e},
       adsurl = {https://ui.adsabs.harvard.edu/abs/2023PSJ.....4..198D}
}

@ARTICLE{2019Icar..328...93D,
       author = {{DeMartini}, Joseph V. and {Richardson}, Derek C. and {Barnouin}, Olivier S. and {Schmerr}, Nicholas C. and {Plescia}, Jeffrey B. and {Scheirich}, Petr and {Pravec}, Petr},
        title = "{Using a discrete element method to investigate seismic response and spin change of 99942 Apophis during its 2029 tidal encounter with Earth}",
      journal = {\icarus},
         year = 2019,
        month = aug,
       volume = {328},
        pages = {93-103},
          doi = {10.1016/j.icarus.2019.03.015},
       adsurl = {https://ui.adsabs.harvard.edu/abs/2019Icar..328...93D}
}

@ARTICLE{2024A&A...682A..93D,
       author = {{\v{D}urech}, J. and {Vokrouhlick{\'y}}, D. and {Pravec}, P. and {Krugly}, Yu. and {Polishook}, D. and {Hanu{\v{s}}}, J. and {Marchis}, F. and {Ro{\.z}ek}, A. and {Snodgrass}, C. and {Alegre}, L. and {Donchev}, Z. and {Ehgamberdiev}, Sh. A. and {Fatka}, P. and {Gaftonyuk}, N.~M. and {Gal{\'a}d}, A. and {Hornoch}, K. and {Inasaridze}, R. Ya. and {Khalouei}, E. and {Ku{\v{c}}{\'a}kov{\'a}}, H. and {Ku{\v{s}}nir{\'a}k}, P. and {Oey}, J. and {Pray}, D.~P. and {Sergeev}, A. and {Slyusarev}, I.},
        title = "{Secular change in the spin states of asteroids due to radiation and gravitation torques. New detections and updates of the YORP effect}",
      journal = {\aap},
         year = 2024,
        month = feb,
       volume = {682},
          eid = {A93},
        pages = {A93},
          doi = {10.1051/0004-6361/202348350},
archivePrefix = {arXiv},
       eprint = {2312.05157},
 primaryClass = {astro-ph.EP},
       adsurl = {https://ui.adsabs.harvard.edu/abs/2024A&A...682A..93D}
}

@INPROCEEDINGS{2022LPICo2681.2007F,
       author = {{Farnocchia}, D. and {Chesley}, S.~R.},
        title = "{Apophis Trajectory, Impact Hazard, and Sensitivity to Spacecraft Contact}",
    booktitle = {Apophis T-7 Years: Knowledge Opportunities for the Science of Planetary Defense},
         year = 2022,
       series = {LPI Contributions},
       volume = {2681},
        month = may,
          eid = {2007},
        pages = {2007},
       adsurl = {https://ui.adsabs.harvard.edu/abs/2022LPICo2681.2007F}
}

@INPROCEEDINGS{2024LPICo3006.2046F,
       author = {{Farnocchia}, D. and {Vokrouhlick{\'y}}, D. and {{\v{C}}apek}, D. and {Chesley}, S.~R. and {DellaGiustina}, D.~N.},
        title = "{The Yarkovsky Effect on Apophis: Bulk Density Constraints and Anticipated OSIRIS-APEX Science}",
    booktitle = {Apophis T-5 Workshop},
         year = 2024,
       series = {LPI Contributions},
       volume = {3006},
        month = apr,
          eid = {2046},
        pages = {2046},
       adsurl = {https://ui.adsabs.harvard.edu/abs/2024LPICo3006.2046F}
}

@ARTICLE{2021Icar..36514493H,
       author = {{Hirabayashi}, M. and {Kim}, Y. and {Brozovi{\'c}}, M.},
        title = "{Finite element modeling to characterize the stress evolution in Asteroid (99942) Apophis during the 2029 Earth encounter}",
      journal = {\icarus},
         year = 2021,
        month = sep,
       volume = {365},
          eid = {114493},
        pages = {114493},
          doi = {10.1016/j.icarus.2021.114493},
       adsurl = {https://ui.adsabs.harvard.edu/abs/2021Icar..36514493H}
}

@ARTICLE{2023MNRAS.520.3405K,
       author = {{Kim}, Yaeji and {DeMartini}, Joseph V. and {Richardson}, Derek C. and {Hirabayashi}, Masatoshi},
        title = "{Tidal resurfacing model for (99942) Apophis during the 2029 close approach with Earth}",
      journal = {\mnras},
         year = 2023,
        month = apr,
       volume = {520},
       number = {3},
        pages = {3405-3415},
          doi = {10.1093/mnras/stad351},
archivePrefix = {arXiv},
       eprint = {2301.13063},
 primaryClass = {astro-ph.EP},
       adsurl = {https://ui.adsabs.harvard.edu/abs/2023MNRAS.520.3405K}
}

@ARTICLE{2024SoSyR..58..208L,
       author = {{Lobanova}, K.~S. and {Melnikov}, A.~V.},
        title = "{Disturbances in the Rotational Dynamics of Asteroid (99942) Apophis at its Approach to the Earth in 2029}",
      journal = {Solar System Research},
         year = 2024,
        month = may,
       volume = {58},
       number = {2},
        pages = {208-219},
          doi = {10.1134/S0038094623700107},
       adsurl = {https://ui.adsabs.harvard.edu/abs/2024SoSyR..58..208L}
}

@ARTICLE{2014A&A...566A..22M,
       author = {{M{\"u}ller}, T.~G. and {Kiss}, C. and {Scheirich}, P. and {Pravec}, P. and {O'Rourke}, L. and {Vilenius}, E. and {Altieri}, B.},
        title = "{Thermal infrared observations of asteroid (99942) Apophis with Herschel}",
      journal = {\aap},
         year = 2014,
        month = jun,
       volume = {566},
          eid = {A22},
        pages = {A22},
          doi = {10.1051/0004-6361/201423841},
archivePrefix = {arXiv},
       eprint = {1404.5847},
 primaryClass = {astro-ph.EP},
       adsurl = {https://ui.adsabs.harvard.edu/abs/2014A&A...566A..22M}
}

@ARTICLE{2022PSJ.....3..123R,
       author = {{Reddy}, Vishnu and {Kelley}, Michael S. and {Dotson}, Jessie and {Farnocchia}, Davide and {Erasmus}, Nicolas and {Polishook}, David and {Masiero}, Joseph and {Benner}, Lance A.~M. and {Bauer}, James and {Alarcon}, Miguel R. and {Balam}, David and {Bamberger}, Daniel and {Bell}, David and {Barnardi}, Fabrizio and {Bressi}, Terry H. and {Brozovic}, Marina and {Brucker}, Melissa J. and {Buzzi}, Luca and {Cano}, Juan and {Cantillo}, David and {Cennamo}, Ramona and {Chastel}, Serge and {Chingis}, Omarov and {Choi}, Young-Jun and {Christensen}, Eric and {Denneau}, Larry and {Dr{\'o}{\.z}d{\.z}}, Marek and {Elenin}, Leonid and {Erece}, Orhan and {Faggioli}, Laura and {Falco}, Carmelo and {Glamazda}, Dmitry and {Graziani}, Filippo and {Heinze}, Aren N. and {Holman}, Matthew J. and {Ivanov}, Alexander and {Jacques}, Cristovao and {Janse van Rensburg}, Petro and {Kaiser}, Galina and {Kami{\'n}ski}, Krzysztof and {Kami{\'n}ska}, Monika K. and {Kaplan}, Murat and {Kim}, Dong-Heun and {Kim}, Myung-Jin and {Kiss}, Csaba and {Kokina}, Tatiana and {Kuznetsov}, Eduard and {Larsen}, Jeffrey A. and {Lee}, Hee-Jae and {Lees}, Robert C. and {de Le{\'o}n}, Julia and {Licandro}, Javier and {Mainzer}, Amy and {Marciniak}, Anna and {Marsset}, Michael and {Mastaler}, Ron A. and {Mathias}, Donovan L. and {McMillan}, Robert S. and {Medeiros}, Hissa and {Micheli}, Marco and {Mokhnatkin}, Artem and {Moon}, Hong-Kyu and {Morate}, David and {Naidu}, Shantanu P. and {Nastasi}, Alessandro and {Novichonok}, Artem and {Og{\l}oza}, Waldemar and {P{\'a}l}, Andr{\'a}s and {P{\'e}rez-Toledo}, Fabricio and {Perminov}, Alexander and {Petrescu}, Elisabeta and {Popescu}, Marcel and {Read}, Mike T. and {Reichart}, Daniel E. and {Reva}, Inna and {Roh}, Dong-Goo and {Rumpf}, Clemens and {Satpathy}, Akash and {Schmalz}, Sergei and {Scotti}, James V. and {Serebryanskiy}, Aleksander and {Serra-Ricart}, Miquel and {Sonbas}, Eda and {Szak{\'a}ts}, Robert and {Taylor}, Patrick A. and {Tonry}, John L. and {Tubbiolo}, Andrew F. and {Veres}, Peter and {Wainscoat}, Richard and {Warner}, Elizabeth and {Weiland}, Henry J. and {Wells}, Guy and {Weryk}, Robert and {Wheeler}, Lorien F. and {Wiebe}, Yulia and {Yim}, Hong-Suh and {{\.Z}ejmo}, Micha{\l} and {Zhornichenko}, Anastasiya and {Zo{\l}a}, Stanis{\l}aw and {Michel}, Patrick},
        title = "{Apophis Planetary Defense Campaign}",
      journal = {\psj},
         year = 2022,
        month = may,
       volume = {3},
       number = {5},
          eid = {123},
        pages = {123},
          doi = {10.3847/PSJ/ac66eb},
       adsurl = {https://ui.adsabs.harvard.edu/abs/2022PSJ.....3..123R}
}

@ARTICLE{2022PSJ.....3..124S,
       author = {{Satpathy}, Akash and {Mainzer}, Amy and {Masiero}, Joseph R. and {Linder}, Tyler and {Cutri}, Roc M. and {Wright}, Edward L. and {Pittichov{\'a}}, Jana and {Grav}, Tommy and {Kramer}, Emily},
        title = "{NEOWISE Observations of the Potentially Hazardous Asteroid (99942) Apophis}",
      journal = {\psj},
         year = 2022,
        month = may,
       volume = {3},
       number = {5},
          eid = {124},
        pages = {124},
          doi = {10.3847/PSJ/ac66d1},
archivePrefix = {arXiv},
       eprint = {2204.05412},
 primaryClass = {astro-ph.EP},
       adsurl = {https://ui.adsabs.harvard.edu/abs/2022PSJ.....3..124S}
}

@ARTICLE{2005Icar..178..281S,
       author = {{Scheeres}, D.~J. and {Benner}, L.~A.~M. and {Ostro}, S.~J. and {Rossi}, A. and {Marzari}, F. and {Washabaugh}, P.},
        title = "{Abrupt alteration of Asteroid 2004 MN4's spin state during its 2029 Earth flyby}",
      journal = {\icarus},
         year = 2005,
        month = nov,
       volume = {178},
       number = {1},
        pages = {281-283},
          doi = {10.1016/j.icarus.2005.06.002},
       adsurl = {https://ui.adsabs.harvard.edu/abs/2005Icar..178..281S}
}

@ARTICLE{2018A&A...617A..74S,
       author = {{Souchay}, J. and {Lhotka}, C. and {Heron}, G. and {Herv{\'e}}, Y. and {Puente}, V. and {Folgueira Lopez}, M.},
        title = "{Changes of spin axis and rate of the asteroid (99942) Apophis during the 2029 close encounter with Earth: A constrained model}",
      journal = {\aap},
         year = 2018,
        month = sep,
       volume = {617},
          eid = {A74},
        pages = {A74},
          doi = {10.1051/0004-6361/201832914},
       adsurl = {https://ui.adsabs.harvard.edu/abs/2018A&A...617A..74S}
}

@article{Santana-Ros2016,
	adsurl = {https://ui.adsabs.harvard.edu/abs/2016MPBu...43..205S},
	author = {{Santana-Ros}, Toni and {Marciniak}, Anna and {Bartczak}, Prezemyslaw},
	journal = {Minor Planet Bulletin},
	month = jul,
	number = {3},
	pages = {205-207},
	title = {{Gaia-GOSA: A Collaborative Service for Asteroid Observers}},
	volume = {43},
	year = 2016}

@article{Jehin2011,
	adsurl = {http://adsabs.harvard.edu/abs/2011Msngr.145....2J},
	author = {{Jehin}, E. and {Gillon}, M. and {Queloz}, D. and {Magain}, P. and {Manfroid}, J. and {Chantry}, V. and {Lendl}, M. and {Hutsem{\'e}kers}, D. and {Udry}, S.},
	journal = {The Messenger},
	month = sep,
	pages = {2-6},
	title = {{TRAPPIST: TRAnsiting Planets and PlanetesImals Small Telescope}},
	volume = 145,
	year = 2011}

@article{Humes2024,
	adsurl = {https://ui.adsabs.harvard.edu/abs/2024PSJ.....5..271H},
	archiveprefix = {arXiv},
	author = {{Humes}, Oriel A. and {Hanu{\v{s}}}, Josef},
	doi = {10.3847/PSJ/ad8f3a},
	eid = {271},
	eprint = {2412.04123},
	journal = {\psj},
	month = dec,
	number = {12},
	pages = {271},
	primaryclass = {astro-ph.EP},
	title = {{Insights on the Rotational State and Shape of Asteroid (203) Pompeja from TESS Photometry}},
	volume = {5},
	year = 2024}

@article{Ricker2015,
	adsurl = {https://ui.adsabs.harvard.edu/abs/2015JATIS...1a4003R},
	author = {{Ricker}, George R. and {Winn}, Joshua N. and {Vanderspek}, Roland and {Latham}, David W. and {Bakos}, G{\'a}sp{\'a}r {\'A}. and {Bean}, Jacob L. and {Berta-Thompson}, Zachory K. and {Brown}, Timothy M. and {Buchhave}, Lars and {Butler}, Nathaniel R. and {Butler}, R. Paul and {Chaplin}, William J. and {Charbonneau}, David and {Christensen-Dalsgaard}, J{\o}rgen and {Clampin}, Mark and {Deming}, Drake and {Doty}, John and {De Lee}, Nathan and {Dressing}, Courtney and {Dunham}, Edward W. and {Endl}, Michael and {Fressin}, Francois and {Ge}, Jian and {Henning}, Thomas and {Holman}, Matthew J. and {Howard}, Andrew W. and {Ida}, Shigeru and {Jenkins}, Jon M. and {Jernigan}, Garrett and {Johnson}, John Asher and {Kaltenegger}, Lisa and {Kawai}, Nobuyuki and {Kjeldsen}, Hans and {Laughlin}, Gregory and {Levine}, Alan M. and {Lin}, Douglas and {Lissauer}, Jack J. and {MacQueen}, Phillip and {Marcy}, Geoffrey and {McCullough}, Peter R. and {Morton}, Timothy D. and {Narita}, Norio and {Paegert}, Martin and {Palle}, Enric and {Pepe}, Francesco and {Pepper}, Joshua and {Quirrenbach}, Andreas and {Rinehart}, Stephen A. and {Sasselov}, Dimitar and {Sato}, Bun'ei and {Seager}, Sara and {Sozzetti}, Alessandro and {Stassun}, Keivan G. and {Sullivan}, Peter and {Szentgyorgyi}, Andrew and {Torres}, Guillermo and {Udry}, Stephane and {Villasenor}, Joel},
	doi = {10.1117/1.JATIS.1.1.014003},
	eid = {014003},
	journal = {Journal of Astronomical Telescopes, Instruments, and Systems},
	month = jan,
	pages = {014003},
	title = {{Transiting Exoplanet Survey Satellite (TESS)}},
	volume = {1},
	year = 2015}

@article{Pal2020,
	adsurl = {https://ui.adsabs.harvard.edu/abs/2020ApJS..247...26P},
	archiveprefix = {arXiv},
	author = {{P{\'a}l}, Andr{\'a}s and {Szak{\'a}ts}, R{\'o}bert and {Kiss}, Csaba and {B{\'o}di}, Attila and {Bogn{\'a}r}, Zs{\'o}fia and {Kalup}, Csilla and {Kiss}, L{\'a}szl{\'o} L. and {Marton}, G{\'a}bor and {Moln{\'a}r}, L{\'a}szl{\'o} and {Plachy}, Emese and {S{\'a}rneczky}, Kriszti{\'a}n and {Szab{\'o}}, Gyula M. and {Szab{\'o}}, R{\'o}bert},
	doi = {10.3847/1538-4365/ab64f0},
	eid = {26},
	eprint = {2001.05822},
	journal = {\apjs},
	month = mar,
	number = {1},
	pages = {26},
	primaryclass = {astro-ph.EP},
	title = {{Solar System Objects Observed with TESS{\textemdash}First Data Release: Bright Main-belt and Trojan Asteroids from the Southern Survey}},
	volume = {247},
	year = 2020}

@MISC{Fausnaugh2021, author = {{Fausnaugh}, Michael M. and {Burke}, Christopher James and {Caldwell}, Douglas A. and {Jenkins}, Jon M. and {Smith}, Jeffrey C. and {Twicken}, Joseph D. and {Vanderspek}, Roland and {Doty}, John P. and {Ting}, Eric B. and {Villasenor}, Joel S.}, title = "{TESS Data Release Notes:Sector 35, DR51}", howpublished = {NASA Technical Memorandum, NASA/STI Accession number: 20210014771}, year = 2021, month = apr, pages = {14771}, adsurl = {https://ui.adsabs.harvard.edu/abs/2021ntrs.rept14771F} }

@article{Polishook2020,
	adsurl = {https://ui.adsabs.harvard.edu/abs/2020Icar..33613415P},
	archiveprefix = {arXiv},
	author = {{Polishook}, David and {Aharonson}, Oded},
	doi = {10.1016/j.icarus.2019.113415},
	eid = {113415},
	eprint = {1904.09627},
	journal = {\icarus},
	month = jan,
	pages = {113415},
	primaryclass = {astro-ph.EP},
	title = {{Surface slopes of asteroid pairs as indicators of mechanical properties and cohesion}},
	volume = {336},
	year = 2020}

@INPROCEEDINGS{Zola2025,
       author = {{Zola}, S. and {Stachowski}, G. and {Kurowski}, S. and {Kundera}, T. and {Waniak}, W. and {Reichart}, D.~E. and {Kouprianov}, V. and {Haislip}, J.~B. and {Wyrzykowski}, L. and {Mikolajczyk}, P.~J.},
        title = "{OAUJ-CDK500: A New Krak{\'o}w Robotic Telescope}",
    booktitle = {Revista Mexicana de Astronomia y Astrofisica Conference Series},
         year = 2025,
       series = {Revista Mexicana de Astronomia y Astrofisica Conference Series},
       volume = {59},
        month = jul,
        pages = {31-36},
          doi = {10.22201/ia.14052059p.2025.59.07},
       adsurl = {https://ui.adsabs.harvard.edu/abs/2025RMxAC..59...31Z}
}

@INPROCEEDINGS{Zola2021,
       author = {{Zola}, S. and {Kouprianov}, V. and {Reichart}, D.~E. and {Bhatta}, G. and {Caton}, D.~B.},
        title = "{Long-term Photometry with Skynet Robotic Telescope Network}",
    booktitle = {Revista Mexicana de Astronomia y Astrofisica Conference Series},
         year = 2021,
       series = {Revista Mexicana de Astronomia y Astrofisica Conference Series},
       volume = {53},
        month = sep,
        pages = {206-214},
          doi = {10.22201/ia.14052059p.2021.53.40},
       adsurl = {https://ui.adsabs.harvard.edu/abs/2021RMxAC..53..206Z}
}

\onecolumn

\begin{appendix}

  \section{Aspect data}

    \begin{longtable}{cccrrrr}
     \caption{\label{tab:aspect_P2021}Aspect data for new observations from 2020--2021 taken with the Danish telescope, ESO, La Silla.}\\
      \hline \hline
        Date    & $r$   & $\Delta$      & $\alpha\phantom{g}$   & \multicolumn{1}{c}{$\lambda$} & \multicolumn{1}{c}{$\beta$}   & \multicolumn{1}{c}{$N$} \\
        & [au]  & [au]          & [deg]                 & \multicolumn{1}{c}{[deg]}         & \multicolumn{1}{c}{[deg]}         & \\
        \hline
      \endfirsthead
      \caption{continued.}\\
      \hline\hline
        Date    & $r$   & $\Delta$      & $\alpha\phantom{g}$   & \multicolumn{1}{c}{$\lambda$} & \multicolumn{1}{c}{$\beta$}   & \multicolumn{1}{c}{$N$} \\
        & [au]  & [au]          & [deg]                 & \multicolumn{1}{c}{[deg]}         & \multicolumn{1}{c}{[deg]}         & \\
      \hline
      \endhead
      \hline
      \endfoot
2020 11 16.3  & 0.915    & 0.263  & 98.6     & 168.2     & $-8.3$  & 7 \\
2020 11 17.3  & 0.918    & 0.264  & 97.6     & 168.4     & $-8.4$  & 10 \\
2020 11 18.3  & 0.922    & 0.265  & 96.7     & 168.6     & $-8.6$  & 7 \\
2020 11 19.3  & 0.925    & 0.266  & 95.8     & 168.8     & $-8.7$  & 6 \\
2020 11 20.3  & 0.929    & 0.267  & 95.0     & 169.0     & $-8.9$  & 6 \\
2020 11 21.3  & 0.932    & 0.268  & 94.1     & 169.2     & $-9.0$  & 6 \\
2020 11 22.3  & 0.935    & 0.268  & 93.3     & 169.5     & $-9.2$  & 6 \\
2020 11 23.3  & 0.939    & 0.269  & 92.4     & 169.7     & $-9.3$  & 7 \\
2020 12 14.3  & 1.004    & 0.265  & 78.2     & 176.0     & $-12.2$  & 7 \\
2020 12 15.3  & 1.007    & 0.264  & 77.6     & 176.4     & $-12.3$  & 11 \\
2020 12 16.3  & 1.009    & 0.263  & 77.0     & 176.7     & $-12.5$  & 13 \\
2020 12 17.3  & 1.012    & 0.262  & 76.4     & 177.0     & $-12.6$  & 20 \\
2020 12 18.3  & 1.015    & 0.261  & 75.8     & 177.3     & $-12.7$  & 20 \\
2020 12 19.3  & 1.017    & 0.260  & 75.3     & 177.6     & $-12.9$  & 18 \\
2020 12 20.3  & 1.020    & 0.258  & 74.7     & 177.9     & $-13.0$  & 7 \\
2020 12 21.3  & 1.023    & 0.257  & 74.1     & 178.2     & $-13.2$  & 8 \\
2020 12 22.3  & 1.025    & 0.256  & 73.5     & 178.5     & $-13.3$  & 13 \\
2020 12 23.3  & 1.028    & 0.254  & 73.0     & 178.8     & $-13.5$  & 21 \\
2021 01 10.3  & 1.066    & 0.218  & 62.3     & 182.9     & $-16.4$  & 14 \\
2021 01 11.3  & 1.067    & 0.216  & 61.7     & 183.0     & $-16.6$  & 12 \\
2021 01 12.3  & 1.069    & 0.214  & 61.1     & 183.2     & $-16.7$  & 16 \\
2021 01 13.3  & 1.071    & 0.211  & 60.4     & 183.3     & $-16.9$  & 14 \\
2021 01 14.3  & 1.072    & 0.209  & 59.7     & 183.3     & $-17.1$  & 20 \\
2021 01 18.3  & 1.078    & 0.199  & 56.9     & 183.5     & $-17.9$  & 20 \\
2021 01 19.3  & 1.079    & 0.196  & 56.2     & 183.4     & $-18.1$  & 15 \\
2021 01 20.3  & 1.081    & 0.193  & 55.4     & 183.4     & $-18.3$  & 16 \\
2021 01 21.2  & 1.082    & 0.191  & 54.7     & 183.3     & $-18.5$  & 19 \\
2021 01 22.3  & 1.083    & 0.188  & 53.9     & 183.3     & $-18.7$  & 16 \\
2021 02 04.2  & 1.095    & 0.155  & 42.2     & 179.6     & $-21.6$  & 34 \\
2021 02 05.2  & 1.096    & 0.153  & 41.2     & 179.1     & $-21.8$  & 19 \\
2021 02 06.2  & 1.096    & 0.150  & 40.1     & 178.5     & $-22.0$  & 34 \\
2021 02 07.1  & 1.097    & 0.148  & 39.1     & 178.0     & $-22.2$  & 12 \\
2021 02 08.2  & 1.097    & 0.146  & 37.9     & 177.2     & $-22.4$  & 43 \\
2021 02 09.1  & 1.098    & 0.143  & 36.9     & 176.6     & $-22.6$  & 19 \\
2021 02 10.2  & 1.098    & 0.141  & 35.7     & 175.8     & $-22.8$  & 16 \\
2021 02 11.3  & 1.098    & 0.139  & 34.5     & 175.0     & $-23.0$  & 20 \\
2021 02 12.3  & 1.099    & 0.137  & 33.4     & 174.1     & $-23.1$  & 24 \\
2021 02 15.2  & 1.099    & 0.131  & 30.0     & 171.3     & $-23.6$  & 26 \\
2021 02 16.1  & 1.099    & 0.129  & 28.9     & 170.4     & $-23.7$  & 33 \\
2021 02 16.3  & 1.099    & 0.129  & 28.7     & 170.2     & $-23.7$  & 21 \\
2021 02 17.3  & 1.099    & 0.127  & 27.7     & 169.2     & $-23.8$  & 27 \\
2021 02 18.1  & 1.099    & 0.126  & 26.8     & 168.2     & $-23.8$  & 10 \\
2021 02 18.3  & 1.099    & 0.126  & 26.5     & 168.0     & $-23.9$  & 7 \\
2021 02 19.1  & 1.099    & 0.124  & 25.7     & 167.1     & $-23.9$  & 13 \\
2021 02 19.4  & 1.099    & 0.124  & 25.5     & 166.8     & $-23.9$  & 13 \\
2021 02 20.1  & 1.099    & 0.123  & 24.8     & 165.9     & $-23.9$  & 20 \\
2021 02 20.3  & 1.099    & 0.123  & 24.6     & 165.6     & $-23.9$  & 23 \\
2021 02 21.1  & 1.099    & 0.121  & 23.9     & 164.6     & $-23.9$  & 35 \\
2021 02 21.3  & 1.099    & 0.121  & 23.7     & 164.3     & $-23.9$  & 13 \\
2021 03 05.3  & 1.093    & 0.113  & 24.5     & 147.8     & $-21.8$  & 21 \\
2021 03 06.2  & 1.092    & 0.113  & 25.6     & 146.4     & $-21.4$  & 20 \\
2021 03 07.1  & 1.092    & 0.113  & 26.6     & 145.2     & $-21.1$  & 21 \\
2021 03 07.3  & 1.092    & 0.113  & 26.9     & 144.9     & $-21.0$  & 7 \\
2021 03 08.1  & 1.091    & 0.113  & 27.9     & 143.9     & $-20.7$  & 17 \\
2021 03 08.2  & 1.091    & 0.113  & 28.1     & 143.7     & $-20.6$  & 19 \\
2021 03 09.1  & 1.090    & 0.113  & 29.2     & 142.5     & $-20.2$  & 27 \\
2021 03 09.3  & 1.090    & 0.113  & 29.5     & 142.3     & $-20.2$  & 13 \\
2021 03 10.1  & 1.089    & 0.113  & 30.6     & 141.3     & $-19.8$  & 19 \\
2021 03 10.3  & 1.089    & 0.114  & 30.9     & 141.0     & $-19.7$  & 12 \\
2021 03 11.0  & 1.088    & 0.114  & 31.9     & 140.1     & $-19.3$  & 6 \\
2021 03 11.2  & 1.088    & 0.114  & 32.2     & 139.8     & $-19.3$  & 19 \\
2021 03 15.1  & 1.084    & 0.117  & 38.0     & 135.4     & $-17.3$  & 21 \\
2021 03 15.2  & 1.083    & 0.117  & 38.3     & 135.2     & $-17.2$  & 12 \\
2021 03 16.1  & 1.082    & 0.117  & 39.5     & 134.3     & $-16.8$  & 16 \\
2021 03 16.2  & 1.082    & 0.118  & 39.8     & 134.1     & $-16.7$  & 18 \\
2021 03 17.1  & 1.081    & 0.118  & 41.1     & 133.3     & $-16.2$  & 21 \\
2021 03 18.1  & 1.080    & 0.119  & 42.6     & 132.3     & $-15.7$  & 26 \\
2021 03 21.1  & 1.076    & 0.123  & 47.0     & 129.6     & $-14.1$  & 35 \\
2021 04 06.1  & 1.047    & 0.143  & 67.5     & 121.2     & $-6.5$  & 15 \\
2021 04 08.1  & 1.042    & 0.146  & 69.8     & 120.6     & $-5.6$  & 12 \\
2021 04 09.0  & 1.040    & 0.147  & 70.9     & 120.3     & $-5.3$  & 8 \\
2021 04 12.1  & 1.033    & 0.151  & 74.1     & 119.6     & $-4.1$  & 14 \\
2021 04 13.0  & 1.031    & 0.152  & 75.2     & 119.4     & $-3.7$  & 12 \\
2021 04 14.0  & 1.028    & 0.153  & 76.3     & 119.2     & $-3.3$  & 11 \\
2021 04 15.0  & 1.026    & 0.154  & 77.3     & 119.1     & $-2.9$  & 10 \\
2021 04 16.0  & 1.023    & 0.155  & 78.3     & 118.9     & $-2.6$  & 10 \\
2021 04 30.0  & 0.984    & 0.166  & 93.2     & 116.9     & $ 2.1$  & 8 \\
2021 05 03.0  & 0.975    & 0.167  & 96.5     & 116.5     & $ 3.0$  & 6 \\
2021 05 06.0  & 0.966    & 0.168  & 100.0     & 115.9     & $ 4.0$  & 7 \\
    \end{longtable}
\noindent{\bf Notes.}   The table lists the time of observation, Apophis's distance from the Sun $r$, from Earth $\Delta$, the solar phase angle $\alpha$, the geocentric ecliptic coordinates $\lambda$ and $\beta$, and the number of photometric data points $N$.  Observations were taken in the R filter and are absolutely calibrated.   
 
    \begin{longtable}{cccrrrrl}
    \caption{\label{tab:aspect} Aspect data for new observations from 2020--2021.}\\
        \hline \hline
        Date    & $r$   & $\Delta$      & $\alpha\phantom{g}$   & \multicolumn{1}{c}{$\lambda$} & \multicolumn{1}{c}{$\beta$}     & \multicolumn{1}{c}{$N$} & Observatory         \\
        & [au]  & [au]          & [deg]                 & \multicolumn{1}{c}{[deg]}         & \multicolumn{1}{c}{[deg]}     &  &      \\
        \hline
\endfirsthead
\caption{continued.}\\
\hline\hline
        Date    & $r$   & $\Delta$      & $\alpha\phantom{g}$   & \multicolumn{1}{c}{$\lambda$} & \multicolumn{1}{c}{$\beta$}     & \multicolumn{1}{c}{$N$} & Observatory         \\
        & [au]  & [au]          & [deg]                 & \multicolumn{1}{c}{[deg]}         & \multicolumn{1}{c}{[deg]}     &  &       \\
\hline
\endhead
\hline
\endfoot
2020 12 25.0  & 1.032    & 0.251  & 72.0     & 179.3     & $-13.7$  & 21     &  Wise, Israel \\               
2021 01 06.0  & 1.058    & 0.228  & 65.0     & 182.2     & $-15.6$  & 52     &  Wise, Israel \\               
2021 01 09.0  & 1.063    & 0.221  & 63.2     & 182.7     & $-16.1$  & 18     &  Wise, Israel \\               
2021 01 10.0  & 1.065    & 0.219  & 62.5     & 182.9     & $-16.3$  & 35     &  Wise, Israel \\               
2021 01 15.3  & 1.074    & 0.206  & 59.0     & 183.4     & $-17.3$  & 48     &  TRAPPIST-South \\             
2021 01 16.3  & 1.075    & 0.204  & 58.3     & 183.4     & $-17.5$  & 56     &  TRAPPIST-South \\             
2021 01 18.4  & 1.078    & 0.198  & 56.8     & 183.5     & $-17.9$  & 13     &  TRAPPIST-South \\             
2021 01 19.3  & 1.079    & 0.196  & 56.2     & 183.4     & $-18.1$  & 62     &  TRAPPIST-South \\             
2021 01 20.2  & 1.081    & 0.194  & 55.5     & 183.4     & $-18.3$  & 48     &  TRAPPIST-North \\             
2021 01 20.3  & 1.081    & 0.193  & 55.4     & 183.4     & $-18.3$  & 61     &  TRAPPIST-South \\             
2021 01 21.3  & 1.082    & 0.191  & 54.6     & 183.3     & $-18.5$  & 62     &  TRAPPIST-South \\             
2021 01 22.2  & 1.083    & 0.188  & 53.9     & 183.3     & $-18.7$  & 59     &  TRAPPIST-North \\             
2021 01 22.2  & 1.083    & 0.188  & 53.9     & 183.3     & $-18.7$  & 42     &  TRAPPIST-South \\             
2021 01 23.2  & 1.084    & 0.186  & 53.2     & 183.2     & $-18.9$  & 65     &  TRAPPIST-North \\             
2021 01 23.2  & 1.084    & 0.186  & 53.1     & 183.2     & $-18.9$  & 81     &  TRAPPIST-South \\             
2021 01 24.2  & 1.086    & 0.183  & 52.3     & 183.0     & $-19.1$  & 57     &  TRAPPIST-North \\             
2021 01 25.2  & 1.087    & 0.181  & 51.5     & 182.9     & $-19.4$  & 46     &  TRAPPIST-North \\             
2021 01 25.3  & 1.087    & 0.180  & 51.4     & 182.9     & $-19.4$  & 28     &  TRAPPIST-South \\             
2021 02 06.2  & 1.096    & 0.150  & 40.1     & 178.5     & $-22.0$  & 55     &  EABA, Argentina \\            
2021 02 07.2  & 1.097    & 0.148  & 39.1     & 177.9     & $-22.2$  & 64     &  EABA, Argentina \\            
2021 02 09.1  & 1.098    & 0.143  & 36.9     & 176.6     & $-22.6$  & 35     &  TRAPPIST-South \\             
2021 02 09.2  & 1.098    & 0.143  & 36.9     & 176.6     & $-22.6$  & 30     &  TRAPPIST-South \\             
2021 02 10.2  & 1.098    & 0.141  & 35.7     & 175.8     & $-22.8$  & 208    &  TRAPPIST-South \\             
2021 02 11.3  & 1.098    & 0.139  & 34.4     & 174.9     & $-23.0$  & 36     &  RBT/PST2 (2x) \\              
2021 02 12.2  & 1.099    & 0.137  & 33.4     & 174.1     & $-23.1$  & 202    &  TRAPPIST-South \\             
2021 02 12.3  & 1.099    & 0.137  & 33.3     & 174.1     & $-23.2$  & 18     &  RBT/PST2 (2x) \\              
2021 02 12.7  & 1.099    & 0.136  & 32.8     & 173.7     & $-23.2$  & 9      &  PROMPT-1, Chile \\            
2021 02 13.2  & 1.099    & 0.135  & 32.2     & 173.2     & $-23.3$  & 198    &  TRAPPIST-South \\             
2021 02 13.3  & 1.099    & 0.135  & 32.2     & 173.2     & $-23.3$  & 154    &  EABA, Argentina \\            
2021 02 14.1  & 1.099    & 0.133  & 31.2     & 172.5     & $-23.4$  & 121    &  TRAPPIST-North \\             
2021 02 14.3  & 1.099    & 0.133  & 31.0     & 172.2     & $-23.5$  & 97     &  TRAPPIST-South \\             
2021 02 15.1  & 1.099    & 0.131  & 30.0     & 171.4     & $-23.6$  & 131    &  EABA, Argentina \\            
2021 02 15.1  & 1.099    & 0.131  & 30.1     & 171.4     & $-23.6$  & 58     &  TRAPPIST-South \\             
2021 02 17.2  & 1.099    & 0.127  & 27.7     & 169.2     & $-23.8$  & 208    &  EABA, Argentina \\            
2021 02 17.4  & 1.099    & 0.127  & 27.5     & 169.0     & $-23.8$  & 12     &  RBT/PST2 \\                   
2021 02 18.4  & 1.099    & 0.126  & 26.5     & 167.9     & $-23.9$  & 29     &  RBT/PST2 \\                   
2021 02 18.7  & 1.099    & 0.125  & 26.2     & 167.6     & $-23.9$  & 9      &  PROMPT-1, Chile (2x) \\       
2021 02 19.0  & 1.099    & 0.125  & 25.9     & 167.2     & $-23.9$  & 36     &  Suhora, Poland \\               
2021 02 19.1  & 1.099    & 0.124  & 25.7     & 167.0     & $-23.9$  & 91     &  TRAPPIST-South \\             
2021 02 19.3  & 1.099    & 0.124  & 25.5     & 166.8     & $-23.9$  & 44     &  RBT/PST2 (2x) \\              
2021 02 21.0  & 1.099    & 0.122  & 23.9     & 164.7     & $-23.9$  & 50     &  Wise, Israel \\               
2021 02 21.0  & 1.099    & 0.122  & 23.9     & 164.7     & $-23.9$  & 12     &  Suhora, Poland \\             
2021 02 21.1  & 1.099    & 0.122  & 23.9     & 164.7     & $-23.9$  & 12     &  TRAPPIST-South \\             
2021 02 21.4  & 1.099    & 0.121  & 23.6     & 164.2     & $-23.9$  & 19     &  RBT/PST2 \\                   
2021 02 22.0  & 1.099    & 0.120  & 23.2     & 163.5     & $-23.9$  & 57     &  Wise, Israel \\               
2021 02 22.7  & 1.098    & 0.120  & 22.7     & 162.6     & $-23.9$  & 14     &  PROMPT-1, Chile (2x) \\       
2021 02 23.1  & 1.098    & 0.119  & 22.4     & 162.1     & $-23.9$  & 34     &  TRAPPIST-South \\             
2021 02 24.0  & 1.098    & 0.118  & 21.9     & 160.8     & $-23.8$  & 142    &  TRAPPIST-North \\             
2021 02 24.1  & 1.098    & 0.118  & 21.8     & 160.6     & $-23.8$  & 249    &  TRAPPIST-South \\             
2021 02 25.0  & 1.098    & 0.117  & 21.4     & 159.5     & $-23.7$  & 32     &  Suhora, Poland \\             
2021 02 25.1  & 1.098    & 0.117  & 21.4     & 159.3     & $-23.7$  & 205    &  TRAPPIST-South \\             
2021 03 02.2  & 1.095    & 0.113  & 22.1     & 152.0     & $-22.7$  & 203    &  TRAPPIST-South \\             
2021 03 03.0  & 1.095    & 0.113  & 22.5     & 151.0     & $-22.5$  & 49     &  Suhora, Poland \\             
2021 03 03.0  & 1.095    & 0.113  & 22.5     & 151.0     & $-22.5$  & 148    &  TRAPPIST-North \\             
2021 03 03.1  & 1.095    & 0.113  & 22.6     & 150.8     & $-22.5$  & 254    &  TRAPPIST-South \\             
2021 03 03.3  & 1.095    & 0.113  & 22.7     & 150.6     & $-22.4$  & 39     &  RBT/PST2 (3x) \\              
2021 03 04.1  & 1.094    & 0.113  & 23.5     & 149.4     & $-22.2$  & 193    &  TRAPPIST-South \\             
2021 03 05.2  & 1.093    & 0.113  & 24.5     & 147.8     & $-21.8$  & 151    &  TRAPPIST-South \\             
2021 03 05.8  & 1.093    & 0.113  & 25.2     & 146.9     & $-21.6$  & 134    &  Wise, Israel \\               
2021 03 06.1  & 1.093    & 0.113  & 25.5     & 146.6     & $-21.5$  & 161    &  TRAPPIST-South \\             
2021 03 06.8  & 1.092    & 0.113  & 26.3     & 145.6     & $-21.2$  & 19     &  CDK, Krakow \\                
2021 03 08.0  & 1.091    & 0.113  & 27.8     & 144.0     & $-20.7$  & 79     &  OMM, Canada \\                
2021 03 08.8  & 1.090    & 0.113  & 28.9     & 142.9     & $-20.3$  & 168    &  Wise, Israel \\               
2021 03 09.1  & 1.090    & 0.113  & 29.3     & 142.5     & $-20.2$  & 280    &  TRAPPIST-South \\             
2021 03 10.0  & 1.089    & 0.113  & 30.5     & 141.4     & $-19.8$  & 17     &  Suhora, Poland \\             
2021 03 10.1  & 1.089    & 0.113  & 30.6     & 141.3     & $-19.8$  & 138    &  EABA, Argentina \\            
2021 03 10.2  & 1.089    & 0.114  & 30.8     & 141.1     & $-19.7$  & 247    &  RBT/PST2 (2x) \\              
2021 03 10.2  & 1.089    & 0.114  & 30.8     & 141.1     & $-19.7$  & 11     &  TRAPPIST-South \\             
2021 03 11.0  & 1.088    & 0.114  & 31.9     & 140.2     & $-19.4$  & 274    &  TRAPPIST-North \\             
2021 03 10.9  & 1.088    & 0.114  & 31.7     & 140.3     & $-19.4$  & 16     &  CDK, Krakow \\                
2021 03 11.0  & 1.088    & 0.114  & 32.0     & 140.1     & $-19.3$  & 13     &  Suhora, Poland \\             
2021 03 11.2  & 1.088    & 0.114  & 32.2     & 139.9     & $-19.3$  & 11     &  EABA, Argentina \\            
2021 03 11.2  & 1.088    & 0.114  & 32.3     & 139.8     & $-19.2$  & 237    &  RBT/PST2 (3x) \\              
2021 03 12.2  & 1.087    & 0.115  & 33.8     & 138.6     & $-18.7$  & 29     &  RBT/PST2 (2x) \\              
2021 03 12.9  & 1.086    & 0.115  & 34.7     & 137.8     & $-18.4$  & 38     &  CDK, Krakow \\                
2021 03 12.9  & 1.086    & 0.115  & 34.8     & 137.8     & $-18.4$  & 70     &  Suhora, Poland \\             
2021 03 13.0  & 1.086    & 0.115  & 34.9     & 137.6     & $-18.4$  & 171    &  TRAPPIST-North \\             
2021 03 14.0  & 1.085    & 0.116  & 36.4     & 136.5     & $-17.9$  & 30     &  TRAPPIST-South \\             
2021 03 14.1  & 1.085    & 0.116  & 36.5     & 136.5     & $-17.8$  & 31     &  TRAPPIST-North \\             
2021 03 14.9  & 1.084    & 0.116  & 37.8     & 135.5     & $-17.4$  & 12     &  CDK, Krakow \\                
2021 03 15.1  & 1.084    & 0.117  & 38.0     & 135.3     & $-17.3$  & 6      &  PROMPT-6, Chile \\            
2021 03 15.0  & 1.084    & 0.117  & 37.9     & 135.4     & $-17.3$  & 27     &  TRAPPIST-South \\             
2021 03 15.9  & 1.083    & 0.117  & 39.3     & 134.5     & $-16.9$  & 37     &  CDK, Krakow \\                
2021 03 16.9  & 1.081    & 0.118  & 40.8     & 133.4     & $-16.3$  & 23     &  Adiyaman, Turkey \\           
2021 03 17.0  & 1.081    & 0.118  & 41.0     & 133.3     & $-16.3$  & 18     &  TRAPPIST-South \\             
2021 03 17.1  & 1.081    & 0.118  & 41.1     & 133.2     & $-16.2$  & 8      &  PROMPT-6, Chile \\            
2021 03 19.8  & 1.078    & 0.121  & 45.1     & 130.7     & $-14.8$  & 82     &  Wise, Israel \\               
2021 03 20.8  & 1.076    & 0.122  & 46.6     & 129.9     & $-14.3$  & 72     &  Wise, Israel \\               
2021 03 22.1  & 1.074    & 0.124  & 48.5     & 128.8     & $-13.6$  & 124    &  TRAPPIST-South \\             
2021 03 25.8  & 1.068    & 0.128  & 53.6     & 126.3     & $-11.7$  & 90     &  Wise, Israel \\               
2021 03 27.1  & 1.066    & 0.130  & 55.3     & 125.5     & $-11.1$  & 233    &  TRAPPIST-South \\             
2021 03 28.1  & 1.064    & 0.131  & 56.6     & 124.9     & $-10.6$  & 228    &  TRAPPIST-South \\             
2021 03 29.1  & 1.063    & 0.133  & 57.9     & 124.4     & $-10.1$  & 237    &  TRAPPIST-South \\             
2021 03 30.0  & 1.061    & 0.134  & 59.1     & 123.9     & $-9.6$   & 139    &  TRAPPIST-South \\             
2021 03 30.1  & 1.061    & 0.134  & 59.2     & 123.9     & $-9.6$   & 175    &  OMM, Canada \\                
2021 03 31.1  & 1.059    & 0.135  & 60.4     & 123.4     & $-9.1$   & 221    &  TRAPPIST-South \\             
2021 04 01.9  & 1.055    & 0.138  & 62.6     & 122.7     & $-8.3$   & 62     &  TRAPPIST-North \\             
2021 04 02.0  & 1.055    & 0.138  & 62.7     & 122.6     & $-8.3$   & 15     &  TRAPPIST-South \\             
2021 04 02.9  & 1.053    & 0.139  & 63.8     & 122.3     & $-7.9$   & 113    &  TRAPPIST-North \\             
2021 04 04.1  & 1.051    & 0.141  & 65.2     & 121.8     & $-7.3$   & 215    &  OMM, Canada \\                
2021 04 04.8  & 1.049    & 0.142  & 66.1     & 121.6     & $-7.0$   & 57     &  Wise, Israel \\               
2021 04 05.0  & 1.049    & 0.142  & 66.3     & 121.5     & $-6.9$   & 41     &  TRAPPIST-South \\             
2021 04 05.8  & 1.047    & 0.143  & 67.2     & 121.2     & $-6.6$   & 45     &  Wise, Israel \\               
2021 04 06.0  & 1.047    & 0.143  & 67.5     & 121.2     & $-6.5$   & 91     &  TRAPPIST-South \\             
2021 04 06.8  & 1.045    & 0.144  & 68.3     & 120.9     & $-6.2$   & 35     &  Wise, Israel \\               
2021 04 07.0  & 1.045    & 0.145  & 68.6     & 120.9     & $-6.1$   & 27     &  TRAPPIST-South \\             
2021 04 07.7  & 1.043    & 0.146  & 69.4     & 120.7     & $-5.8$   & 20     &  Wise, Israel \\               
2021 04 09.1  & 1.040    & 0.147  & 70.9     & 120.3     & $-5.2$   & 123    &  TRAPPIST-South \\             
2021 04 09.1  & 1.040    & 0.147  & 70.9     & 120.3     & $-5.2$   & 140    &  OMM, Canada \\                
2021 04 11.0  & 1.035    & 0.150  & 73.1     & 119.8     & $-4.5$   & 84     &  TRAPPIST-South \\             
2021 04 12.0  & 1.033    & 0.151  & 74.1     & 119.6     & $-4.1$   & 93     &  TRAPPIST-South \\             
2021 04 13.1  & 1.031    & 0.152  & 75.2     & 119.4     & $-3.7$   & 116    &  OMM, Canada \\                
   \end{longtable}
\noindent{\bf Notes.} The table lists the time of observation, Apophis's distance from the Sun $r$, from Earth $\Delta$, the solar phase angle $\alpha$, the geocentric ecliptic coordinates $\lambda$ and $\beta$, the number of photometric data points $N$, and the name of the observatory.

\newpage
      \section{Light curve fit}

      \begin{figure*}[h]
        \includegraphics[width=\textwidth]{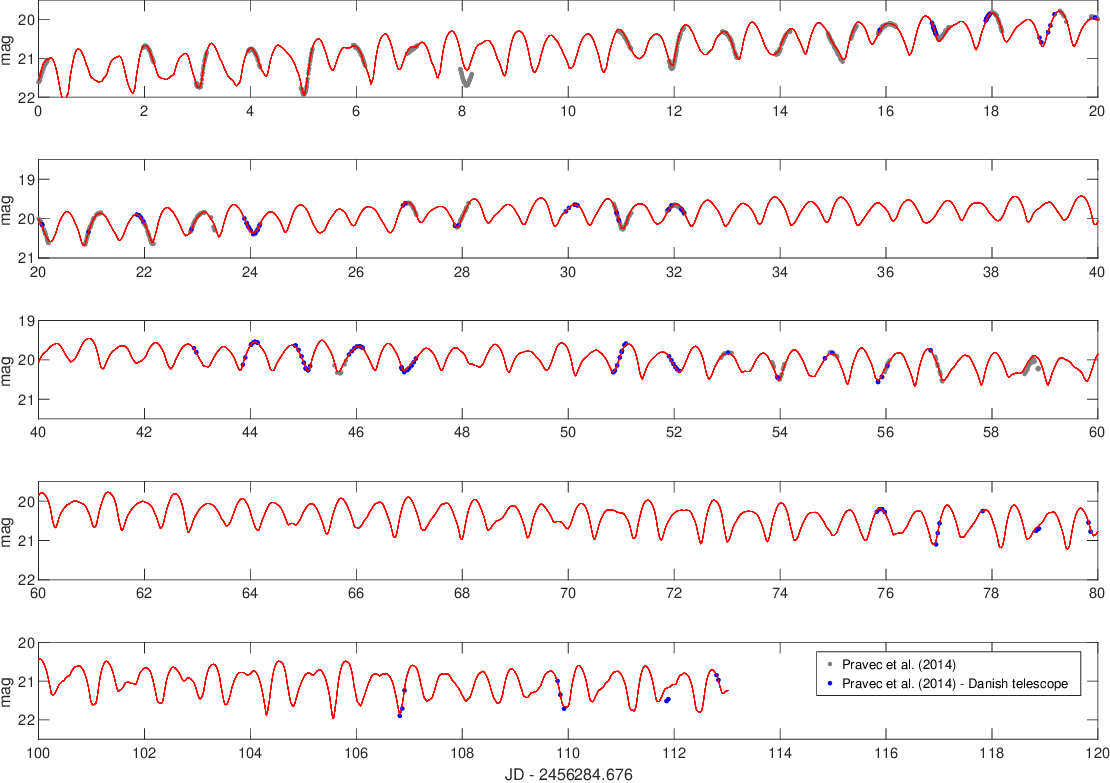}    
        \caption{Fit of the model to the photometric data from 2012--2013. The red curve is the synthetic light curve produced by model A*. The points are photometric measurements published by \cite{Pra.ea:14}. The blue color marks the accurately calibrated subset observed with the Danish telescope, which was used to reconstruct model A. Some of the grey light curves, which were taken with other telescopes, are shifted with respect to the model due to their imperfect photometric calibrations. The vertical axis is the Cousins R reduced to the unit distances from Earth and Sun.}
        \label{fig:lc_fit_2013}
    \end{figure*}

    \begin{figure*}[t]
        \centering\includegraphics[width=0.95\textwidth]{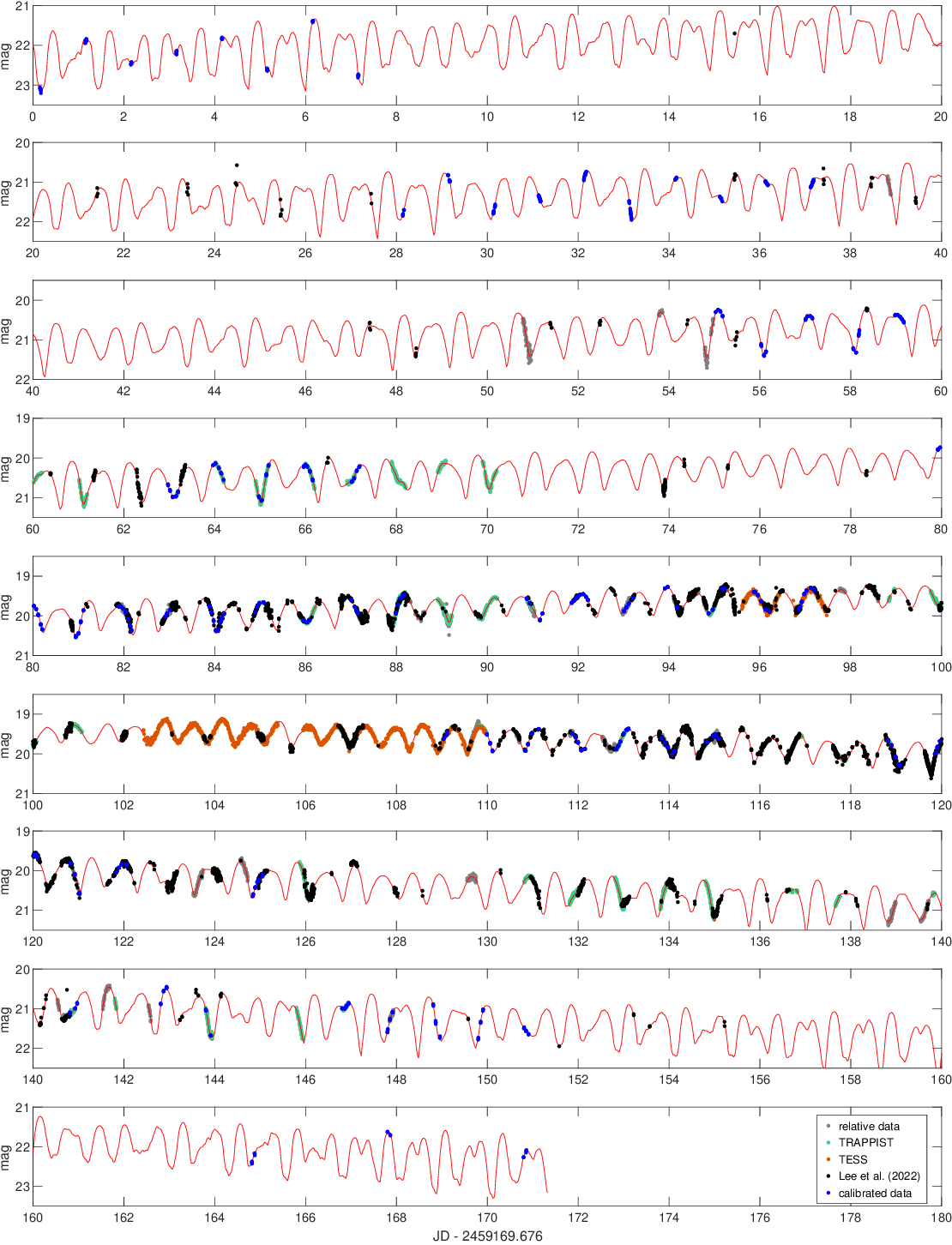}    
        \caption{Fit of the model to the photometric data from 2020--2021. The red curve is the synthetic light curve produced by model A*. The points are photometric measurements from various sources (see the legend and Table~\ref{tab:aspect}). The blue points represent absolutely calibrated photometry from the Danish telescope that was used to reconstruct model A. The vertical axis is the Cousins R reduced to the unit distances from Earth and Sun.}
        \label{fig:lc_fit_2021}
    \end{figure*}

\end{appendix}

\end{document}